\documentclass[showpacs,prd,amsfonts,amsmath,twocolumn,aps,preprintnumbers,letterpaper]{revtex4-1}
\usepackage{amsmath,amssymb}
\usepackage{float}
\usepackage{slashed}
\usepackage{graphicx}
\usepackage{epsfig}
\usepackage{bm}
\usepackage[top=1.0in,bottom=1.0in,left=1.0in,right=1.0in]{geometry}
\usepackage[caption=false]{subfig}
\usepackage[pdftex, citecolor=blue, urlcolor=blue, linkcolor=blue, colorlinks=true, bookmarksopen=true]{hyperref}
\usepackage{tikz}
\usetikzlibrary{positioning,arrows,fit}
\usetikzlibrary{decorations.pathmorphing}
\usetikzlibrary{decorations.markings}
\tikzset{
fermion/.style={thick,draw=black, postaction={decorate},
    decoration={markings,mark=at position .5 with {\arrow[black]{triangle 45}}}},
pomeron/.style={thick,decorate,draw=black,
    decoration={coil,aspect=0}},
Odderon/.style={thick,decorate, draw=black,
    decoration={zigzag,aspect=0}}
 }

\newcommand{\be}{\begin{equation}}
\newcommand{\ee}{\end{equation}}
\newcommand{\bear}{\begin{eqnarray}}
\newcommand{\eear}{\end{eqnarray}}
\newcommand{\ba}{\begin{array}}
\newcommand{\ea}{\end{array}}

\def\be{\begin{eqnarray}}
\def\ee{\end{eqnarray}}
\def\bea{\be}
\def\eea{\ee}

\def\roughly#1{\mathrel{\raise.3ex\hbox{$#1$\kern-.75em%
\lower1ex\hbox{$\sim$}}}}

  \long\def\comment#1{ }

  \newcommand{\beq}{\begin{eqnarray}}
  \newcommand{\eeq}{\end{eqnarray}}
  
 \def\simge{\mathrel{%
   \rlap{\raise 0.511ex \hbox{$>$}}{\lower 0.511ex \hbox{$\sim$}}}}
\def\simle{\mathrel{
   \rlap{\raise 0.511ex \hbox{$<$}}{\lower 0.511ex \hbox{$\sim$}}}}

\renewcommand{\Im}{{\rm Im\,}}
\renewcommand{\Re}{{\rm Re\,}}

\def\D0{D$\slashed{\mathrm{O}}$}

\begin{document}


\title{Holographic Odderon at TOTEM\,?}

\author{Florian Hechenberger}
\email{florian.hechenberger@tuwien.ac.at}
\affiliation{Institut fur Theoretische Physik, Technische Universitat Wien, Wiedner Hauptstrasse 8-10, A-1040 Vienna, Austria
}

\author{Kiminad A. Mamo}
\email{kamamo@wm.edu}
\affiliation{
Physics Department, William and Mary, Williamsburg, VA 23187, USA}

\author{Ismail Zahed}
\email{ismail.zahed@stonybrook.edu}
\affiliation{Center for Nuclear Theory, Department of Physics and Astronomy, Stony Brook University, Stony Brook, New York 11794-3800, USA}

\date{\today}

\begin{abstract}
We consider the contribution of the Odderon to diffractive $pp$ and $p\bar p$ elastic scattering at large center of mass energy. We identify the Odderon and Pomeron with the Reggeized 
$1^{\pm-}$ and $2^{++}$ glueballs in the bulk, respectively. We use for the gravity dual description the repulsive wall model,  to account for the proper Gribov diffusion for off-forward scattering. The eikonalized and unitarized 
amplitudes exhibit a vanishingly small rho-parameter, and a slope parameter
fixed by twice the closed string slope. 
The results for the differential and total cross sections are compared to the empirical results reported recently by the TOTEM collaboration. 
\end{abstract}

\maketitle

%

\section{Introduction}
\label{Introduction}
In the Regge limit, diffractive $pp$  and $p\bar p$ scattering at large $\sqrt{s}$ is dominated by the Pomeron,
with a smaller admixture from 
the Odderon~\cite{Braun:1998fs}, 
a tower of C-odd soft gluons. The Odderon is believed to be the C-odd partner of the C-even Pomeron. While the latter in its
soft version dominates the diffractive $pp$ cross section at high energy, the manifestation of the former is still being debated, 
although recent results from the TOTEM collaboration have claimed it~\cite{TOTEM:2020zzr}.

Regge theory predicts the rise of the $pp$ and $p\bar p$ elastic cross sections at collider energies. The rise follows from 
Regge poles. The dominant pole is the Pomeron with the largest intercept and positive signature. The Odderon carries a
smaller intercept and negative signature. Negative signature  Reggeons add in the $pp$ channel, and subtract in the $p\bar p$ channel.

The hard QCD Pomeron at weak coupling, is a Reggeized BFKL ladder which resums the rapidity ordered collinear emissions. It is a C-even and P-even gluon ladder,
which is identified with j-plane branch-points in the conformal limit. In the same regime, the Odderon is a Reggeized BKP ladder~\cite{Bartels:1980pe,Kwiecinski:1980wb}. It is a C-odd and P-odd exchange of gluons, with two nearby j-plane branch points. At strong coupling, in dual gravity, the Pomeron is identified with a Reggeized spin-j graviton,
while the Odderon with a Reggeized spin-j Kalb-Ramond field~\cite{Brower:2008cy}.

The purpose of this work is to explore the possible contribution of the Odderon in diffractive $pp$ and $p \bar p$ elastic scattering, in light of the recently reported Odderon 
by the \D0 and TOTEM collaborations~\cite{TOTEM:2020zzr}.
In gravity dual formulations, the general aspects of the  Odderon were initially discussed in~\cite{Brower:2008cy}. Our approach
will use the gravity dual formulation, with a repulsive wall to account for confinement. Only in this case, a full Reggeization
with the famed Gribov diffusion is realized.
 For completeness, we note that the Odderon contribution 
in $pp$ and $p\bar p$ scattering, was partially analyzed using an effective string theory in~\cite{Kharzeev:2017azf}, and  the AdS/CFT in the conformal limit in~\cite{Avsar:2009hc}. 

The outline of the paper is as follows: in section~\ref{SEC-II}
we briefly review the salient features of the newly released
$pp$ and $p\bar p$ data by the TOTEM collaboration. In section~\ref{SEC-III} we will 
formulate the gravity dual description using the repulsive wall
model. We will explicitly construct the bulk-to-bulk Pomeron and Odderon propagators, and show how the Reggeized forms of the Pomeron and Odderon emerge from pertinent 
analytical resummations. Unlike the soft wall, the repulsive wall reproduces the expected Gribov diffusion for the Reggeons on the boundary in the confining limit.
The logarithmic diffusion in the conformal limit is shown to hold. In section~\ref{SEC-IV}
we derive the holographic and Reggeized amplitudes for elastic $pp$ and $p\bar p$ scattering
with the associated total and differential cross sections. The rho
and slope parameters are made explicit. In section~\ref{SEC_EIK} the holographic amplitudes are eikonalized. The Froissart bound is recovered 
at asymptotic energies. The eikonal results for the 
slope and rho parameters, as well as the elastic
differential cross section are compared to the recently reported TOTEM data. Our conclusions are in section~\ref{SEC-VI}. More details regarding some aspects of the repulsive and soft wall as well as the conformal limit are given in a number of additional Appendices.

\section{\texorpdfstring{Odderon in $pp$ and $p\bar p$}{Odderon in pp and ppbar}}
\label{SEC-II}
\begin{figure}[!htb]
    \begin{minipage}{.25\textwidth}
       \raggedleft
        \begin{tikzpicture}[node distance=1cm and 1.5cm]
            \coordinate[label={[yshift=4pt]left:$p$}] (e1);
            \coordinate[right=of e1] (aux1);
            \coordinate[right=of aux1,label={[yshift=4pt]right:$p$}] (e2);
            \coordinate[below=2cm of aux1] (aux2);
            \coordinate[left=of aux2,label={[yshift=-4pt]left:$p$}] (e3);
            \coordinate[right=of aux2,label={[yshift=-4pt]right:$p$}] (e4);
            
            \draw[fermion] (e1) -- (aux1);
            \draw[fermion] (aux1) -- (e2);
            \draw[fermion] (e3) -- (aux2);
            \draw[fermion] (aux2) -- (e4);
            \draw[pomeron,double] (aux1) -- node[label={right:$\mathbb{P}$}] {} (aux2);
        \end{tikzpicture}
        \centering (a)
        \label{fig:pomeronEx}
    \end{minipage}%
    \begin{minipage}{0.25\textwidth}
    \raggedright
        \begin{tikzpicture}[node distance=1cm and 1.5cm]
            \coordinate[label={[yshift=4pt]left:$p$}] (e1);
            \coordinate[right=of e1] (aux1);
            \coordinate[right=of aux1,label={[yshift=4pt]right:$p$}] (e2);
            \coordinate[below=2cm of aux1] (aux2);
            \coordinate[left=of aux2,label={[yshift=-4pt]left:$p$}] (e3);
            \coordinate[right=of aux2,label={[yshift=-4pt]right:$p$}] (e4);
            
            \draw[fermion] (e1) -- (aux1);
            \draw[fermion] (aux1) -- (e2);
            \draw[fermion] (e3) -- (aux2);
            \draw[fermion] (aux2) -- (e4);
            \draw[Odderon] (aux1) -- node[label={right:$\mathbb{O}$}] {} (aux2);
        \end{tikzpicture}
        \centering (b)
        \label{fig:OdderonEx}  
   \end{minipage}
   \caption{Feynman diagrams for diffractive $pp$ eleastic scattering through (a) Pomeron and (b) Odderon exchange}
   \label{fig:PO}
\end{figure}
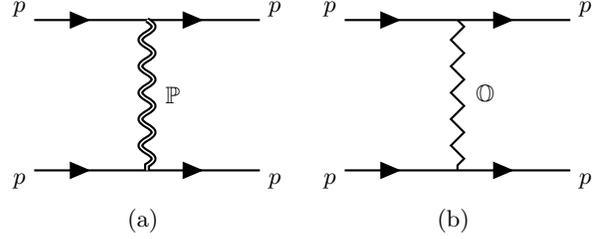

\begin{figure}[!htb]
\includegraphics[height=5cm,width=8cm]{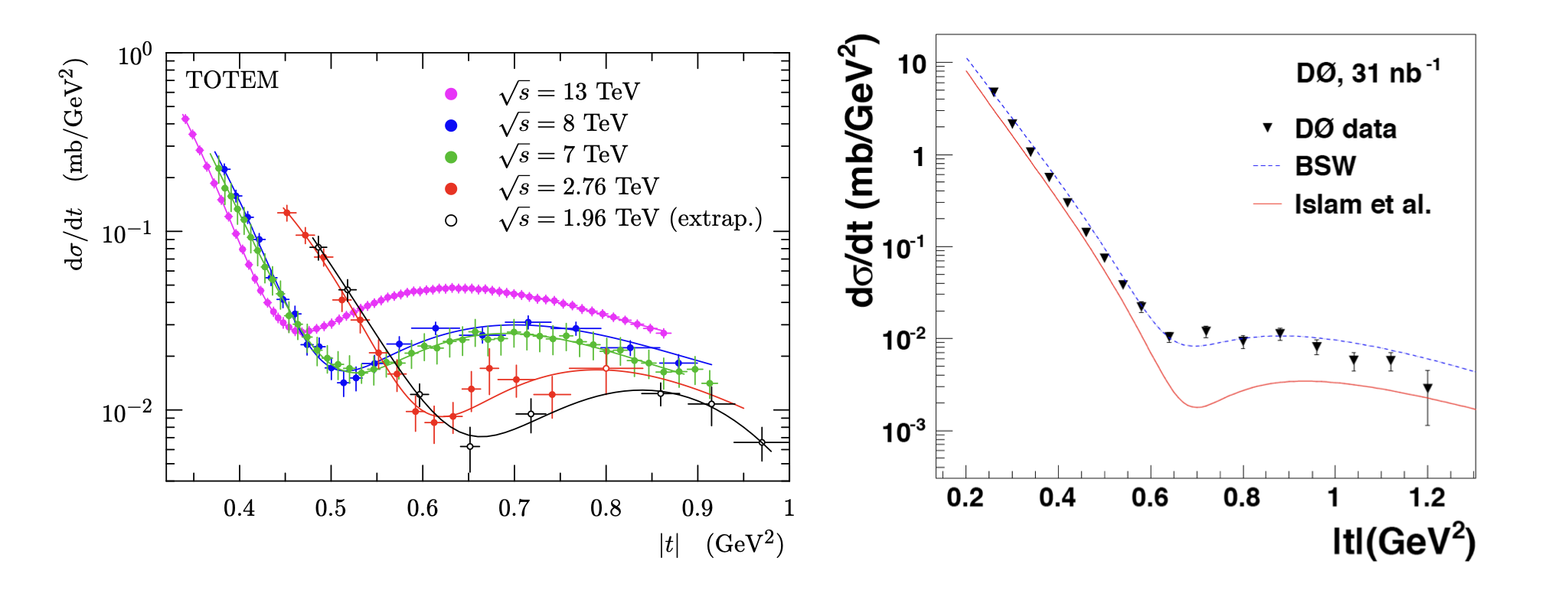}
  \caption{$pp$ differential cross sections from the TOTEM collaboration (left), and $p\bar p$ from the \D0 collaboration (right), for different center of mass energies $\sqrt{s}$~\cite{TOTEM:2020zzr,Royon:2022vyh}.}
  \label{fig:TODO}
\end{figure}

Diffractive $pp$ and $p\bar p$ elastic scattering is dominated by Reggeized glueball exchanges as illustrated  in Fig.~\ref{fig:PO}.
The Pomeron and Odderon carry positive and negative C-parity, respectively. In pQCD the Pomeron is a Reggeized exchange of a two-gluon ladder with rung induced by Lipatov vertices with positive parity $2^{++}$, and the Odderon a Reggeized three-gluon ladder with negative parity $1^{--}$.
Their contribution adds in the elastic $pp$ scattering amplitude, and subtracts in $p\bar p$, 
\bea
\label{APPPPBAR}
{\cal A}_{pp}&=&{\cal A}_{pp}^\mathbb{P}+{\cal A}_{pp}^\mathbb{O}\nonumber\\
{\cal A}_{p\bar p}&=&{\cal A}_{pp}^\mathbb{P}-{\cal A}_{pp}^\mathbb{O}.
\eea
The difference between the two amplitudes at large $\sqrt{s}$ stems from the Odderon exchange, as it discriminates $p$ from $\bar p$ by charge conjugation. The combined and large $\sqrt{s}$ elastic $pp$ data from LHC and $p\bar p$ data from \D0,  were recently analyzed by the TOTEM collaboration, with the results shown in Fig.~\ref{fig:TODO}.  

The data for the differential cross section shows a clear diffractive dip and bump pattern for the $pp$ channel, with a large bump-to-dip ratio that appears to be decreasing for larger $\sqrt{s}$. In the $p\bar p$ scattering data it appears to be flat already at comparatively low $\sqrt{s}$. Among the new and unexpected features of the elastic amplitude are: 1/ the  persistent diffractive structure at low-t; 2/ the absence of additional secondary structures at large-t; 3/ the linear rise of the slope parameter.
We now proceed to analyze \eqref{APPPPBAR} and their pertinent differential and total cross sections in the context of dual gravity. 

\section{Gravity dual description}
\label{SEC-III}
In the QCD dual gravity approach, the precursors to the Pomeron and the Odderon are identified with the graviton $2^{++}$ and the Kalb-Ramond $1^{\pm -}$ fields, each to leading order sourced by the QCD boundary operators
\bea
\label{OP1X}
h^{\mu\nu}[2^{++}]\,:\,\, &&G^{a\mu\alpha}G^{a\nu}_\alpha\nonumber\\
B^{\mu\nu}[1^{+-}]\,:\,\,  && d^{abc}G^{a\alpha\beta}G^{b}_{\alpha\beta}{G}^{c\mu\nu}\nonumber\\
C^{\mu\nu}[1^{--}]\,:\,\,  && d^{abc}G^{a\alpha\beta}G^{b}_{\alpha\beta}{\tilde G}^{c\mu\nu}
\eea
with all traces subtracted. The even spin-j trajectory for the Pomeron, and the odd spin-j trajectory for the Odderon are sourced, respectively, by the boundary operators
\be
\label{OP2X}
h_j^{\mu\nu}[(2+j)^{++}]\,:\,\, &&
G^{a\mu\alpha}D_{\alpha_1}...D_{\alpha_j}G^{a\nu}_\alpha\nonumber\\
B_j^{\mu\nu}[(1+j)^{+-}]\,:\,\,  &&d^{abc}G^{a\alpha\beta}
D_{\alpha_1}...D_{\alpha_j}
G^{b}_{\alpha\beta}{G}^{c\mu\nu}\nonumber\\
C_j^{\mu\nu}[(1+j)^{--}]\,:\,\,  &&d^{abc}G^{a\alpha\beta}
D_{\alpha_1}...D_{\alpha_j}
G^{b}_{\alpha\beta}{\tilde G}^{c\mu\nu}\nonumber\\
\ee
with the proper symmetrization assumed. In particular, 
$h_j$ has conformal dimension $\Delta_h=4+j$ and twist $\tau_h=2+j$,
and {$B_j,\ C_j$} have conformal dimension $\Delta_h=6+j$ and twist $\tau_h=5+j$, respectively. 



\subsection{Dual gravity}
We now proceed to analyse the gravity dual description of the corresponding bulk fields, using the bottom-up approach with a  repulsive wall.
As we detail in Appendix~\ref{ap:b2bSW}, the soft wall model fails to capture Gribov diffusive behaviour, necessary for off-forward scattering. We will thus take a different path and linearize the bulk equations of motion in the dilaton, which resembles an additional potential term. 
To this end we write the metric as
\bea
\label{eq:dsrw}
ds^2&=&e^{2A(z)}(dz^2+\eta_{\mu\nu}dx^\mu dx^\nu), \nonumber\\
e^{2A(z)}&=&\left(\frac{R}{z}\right)^2e^{a \kappa^2 z^2}
\eea
which is amenable to passing from the string frame to Einstein frame. We also introduced a constant prefactor $a$, to study the effects of a repulsive versus attractive dilaton, as well as the open versus closed string coupling. The repulsive dilaton
bears much in common with the Nambu-Goto analysis of the Pomeron  with a hard wall~\cite{Liu:2018gae} (and references therein).
Matching the parameters of type IIB supergravity (SUGRA) on $\rm{AdS}_5$ to  ${\cal N}=4$ Super Yang–Mills  we get
\be
R^2=\sqrt{\lambda} \alpha'.
\ee
In the following we will set $R=1$.


\subsection{Reggeized even spin-j}
The starting point for the Reggeization of the even spin-j exchange is the graviton with spin $j=2$. We recall that the bulk equation of motion for the graviton follows from the linearization of the Einstein-Hilbert action. The traceless and transverse part of the metric fluctuation in physical gauge $\epsilon_{\mu\nu}^{TT}h_{j=2}$, is amenable to
a scalar equation with anomalous dimension 
\bea
\label{DQ1P}
\Delta_g(j=2)=2+\sqrt{4+m_5^2R^2}
\eea
with $m_5^2R^2=0$.
The closed string exchange resumming the even  spin-j $h_j$ fields, can be sought as a graviton exchange with the anomalous dimension~\cite{Mamo:2022jhp}
\bea
\label{DQ1PAN1}
\Delta_g(j)=2+\sqrt{4+m_5^2R^2+m_j^2R^2}
\eea
deformed by the closed  string quantized mass spectrum
\be
j=2+\frac 12 \alpha^\prime m_j^2.
\ee
In other words, the anomalous dimension (\ref{DQ1PAN1}) is
\bea
\Delta_g(j)=2+\sqrt{2\sqrt\lambda(j-j_{\mathbb P})}
\eea
with
\bea
j_{\mathbb P}=2-\frac{4}{2\sqrt\lambda}.
\label{eq:jPconf}
\eea
Higher order corrections in $1/\sqrt{\lambda}$ to the Pomeron intercept, have been analyzed in~\cite{Brower:2014wha}.

We will seek the Reggeized bulk-to-bulk graviton propagator in terms of its Mellin transform 
with the even-signature contribution projected out via the Sommerfeld-Watson transform
\bea
\label{SWT+}
&&G_2(s,t,z,z')=\nonumber\\
&&\int \frac{dj}{4\pi i}\,\frac{(\alpha' s)^{j-2}+(-\alpha' s)^{j-2}}{\sin\pi(j-2)}\,G_2(j,t,z,z')\nonumber\\
\eea
and the analytical continuation in spin $j=2\rightarrow j$ assumed. 

\subsection{Reggeized odd spin-j}
Similarly to the Reggeized even spin-j exchange, the starting point for the Reggeization of the odd   spin-j exchanges
that sums up to a closed string exchange, is the spin-1  $C_2\sim\star dC$
gauge field.  For that, we first recall that the bulk equation of motion for the rescaled gauge field $zC$, is the same as that of a scalar field with anomalous dimension~\cite{Mamo:2019mka}
\bea
\label{DQ1}
\Delta_g(j=1)=2+\sqrt{4+m_5^2R^2}
\eea
with $$m_5^2R^2=-4+m_k^2$$ 
where we added $m_k^2=k^2, (4+k)^2$, for the two branches of the Odderon~\cite{Brower:2008cy}. The latter branch identifies with the canonical dimensions of (\ref{OP1X}-\ref{OP2X}). The former branch is more sensitive to the presence of the adjoint scalars in the SUSY version. 
In this spirit, 
the closed string exchange resumming the odd spin-j $C_j$ gauge fields, can be sought as a vector exchange with the 
anomalous dimension
\bea
\label{AN1}
\Delta_g(j)=2+\sqrt{4+m_5^2R^2+\tilde m_j^2R^2}
\eea
deformed by the closed  string quantized mass spectrum
\be
j=1+\frac 12 \alpha^\prime \tilde m_j^2.
\ee
In other words, the anomalous dimension (\ref{AN1}) is
\bea
\Delta_g(j)=2+\sqrt{2\sqrt\lambda(j-j_{\mathbb O})}
\eea
with
\bea
j_{\mathbb O}=1-\frac{m_k^2}{2\sqrt\lambda}.
\label{eq:jOconf}
\eea
Higher order corrections in $1/\sqrt{\lambda}$ to the Odderon intercept, have been analyzed in~\cite{Brower:2014wha}.
Similarly as for the graviton, the Reggeized Kalb-Ramond field is given by the odd-signature Sommerfeld-Watson transform
\bea
\label{SWT-}
&&G_1(s,t,z,z')=\nonumber\\
&&\int \frac{dj}{4\pi i}\,\frac{(\alpha' s)^{j-1}+(-\alpha' s)^{j-1}}{\sin\pi(j-1)}\,G_1(j,t,z,z').\nonumber\\
\eea   
To conclude, the Reggeized bulk-to-bulk propagators that resum the even and odd trajectory are given by
\begin{widetext}
  \bea
\label{SWT}
G_{j_\pm}(s,t,z,z')=\int \frac{dj}{4\pi i}\,\frac{(\alpha' s)^{j-j_\pm}+(-\alpha' s)^{j-j_\pm}}{\sin\pi(j-j_\pm)}\,G_{j_\pm}(j,t,z,z'),\qquad  j_+=2, \qquad j_-=1
\eea  
\end{widetext}

\subsection{Resummed bulk-to-bulk propagator}
The analytically continued spin-2 and spin-0 are related by
\bea
G_2(j,t,z,z')=e^{-(j-2)(A(z)+A(z'))}\,G_0(j,t,z,z')\nonumber\\
\eea
with the warped scalar propagator obtained from
\bea
\label{eq:SL}
L_zG_0(z,z')=\frac {\delta(z-z')}{w(z)}
\eea
and the pertinent Sturm-Liouville operator
\bea
\label{eq:SLX}
L_z=\frac 1{w(z)}\,d_z(w(x)\,p_0(z)\,d_z)+p_2(z).
\eea
For the background \eqref{eq:dsrw} the scalar bulk-to-bulk propagator obeys the Sturm-Liouville 
equation (\ref{eq:SL}-\ref{eq:SLX}) with weights
\bea
w(z)&=&\sqrt{g}\nonumber\\
p_0(z)&=&-g^{zz}(z)\nonumber\\
p_2(z)&=&S_j-tz^2,
\eea
where $$S_j=m_5^2R^2+m_j^2R^2\,.$$
After carrying out the coordinate transformations and field redefinitions laid out in Appendix~\ref{ap:b2bSW}, and keeping in mind that the dilaton has been absorbed into the metric, the scalar bulk-to-bulk propagator is now seen to solve
\begin{widetext}
\bea
\label{WHIT:lin}
K_0''(v)+\left(-\frac{S_j(1+\frac{2}{3}v)}{4v^2}+\frac{v(2-v)-3}{4v^2}+\frac{t/a}{2v}\right)K_0(v)=\frac{\delta(v-v')}{\sqrt{6}a\kappa}
\eea
\end{widetext}
where we expanded the dilaton to linear order $e^{2av/3}=1+2av/3$. The independent homogeneous solutions to \eqref{WHIT:lin} are Whittaker functions
\bea
\label{WHIT}
K_1(v)&=&e^{-\frac v2}v^{\frac 12+\alpha}\,
\mathbb M\bigg(\frac 12 +\alpha-\beta, 1+2\alpha, v\bigg)\nonumber\\
K_2(v)&=&e^{-\frac v2}v^{\frac 12+\alpha}\,
\mathbb U\bigg(\frac 12 +\alpha-\beta, 1+2\alpha, v\bigg)\nonumber\\
\eea
with Kummer  $\mathbb M$ (regular at $v=0$)  and Tricomi $\mathbb U$ (irregular with branch cut at $v=0$) hypergeometric functions and

\be
\alpha=\frac{\Delta_g(j)-2}{2},\ \beta=\frac{3-S_j-m_5^2+3\tilde t/a}{6}.
\ee
The inhomogeneous solution to \eqref{WHIT:lin} is then

\bea
\label{PROP1lin}
K_0(v, v')&=&\frac 12 {\cal A}\, K_2(v)K_1(v')\qquad v>v'\nonumber\\
K_0(v, v')&=&\frac 12 {\cal A}\, K_1(v)K_2(v')\qquad v<v'\nonumber\\
\eea
with the normalization fixed by the Wronskian

\bea
{\cal A}^{-1}=-\sqrt{6}a\kappa{\cal W}(K_2,K_1)=
-\frac{\sqrt{6}a\kappa\Gamma(1+2\alpha)}{\Gamma\bigg(\frac 12 +\alpha-\beta\bigg)}.
\nonumber\\
\eea
The cost of linearizing the dilaton is a shift in the conformal intercept resulting from the poles of the gamma function given by
\be
\label{evenTrs}
j_{\mathbb{P}}(t)=j_+-\frac{3}{2\sqrt{\lambda}}+\frac{\alpha'}{2}t
\ee
under the condition that $5a\kappa^2 R^2=2$. From this identification it is apparent that only the repulsive dilaton with $a>0$ gives the correct diffusive Regge behaviour, since the attractive dilaton with $a<0$ would violate unitarity. As shown below and in Appendix~\ref{ap:b2bSW}, this choice shifts the intercept slightly below the result from the soft wall or pure AdS, which is however to be expected since linearizing the dilaton introduces an additional potential term. We also find a pole with a higher intercept

\be
j_\mathbb{P}(t)=j_++\frac{6}{\sqrt{\lambda}}+\frac{\alpha'}{2}t
\ee
coming from the positive root which, however, will not be picked up by the contour integration of the Sommerfeld-Watson transform.

\def\xr{4}
\def\yr{1}
\providecommand{\poles}{
  \node (poles) at (2.5,1.5){};
  \draw[fill]
  (1,0) coordinate [circle,fill,inner sep=1.25pt] (p1)
  (2,0) coordinate [circle,fill,inner sep=1.25pt] (p2)
  (3,0) coordinate [circle,fill,inner sep=1.25pt] (p3)
  (.45,0) coordinate [circle,fill,inner sep=1.25pt] (p4);
}
\providecommand{\polecontours}{
  \draw[blue!60!black,decoration={markings,mark=between positions 0.03 and 1.03 step 0.12 with \arrow{<}},postaction={decorate}] (p1) circle (0.25) node [below=0.3] {$1$} (p2) circle (0.25) node [below=0.3] {$3$} (p3) circle (0.25) node [below=0.3] {$5$} (p4) node [below=0.3] {$j_{\mathbb O}$};
}

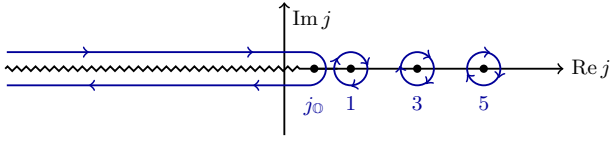
\begin{figure}[!htb]
\resizebox{.5\textwidth}{!}{
    \centering
    \begin{tikzpicture}[thick]
  \draw [decorate,decoration={zigzag,segment length=4,amplitude=1,post=lineto,post length=15}] (-1.05*\xr,0) -- (0.8,0);
  \draw [->,decorate,decoration={segment length=4,amplitude=1,pre=lineto,pre length=15,post=lineto,post length=5}] (.5,0) -- (1.05*\xr,0) node [right] {$\Re j$};
  \draw [->] (0,-\yr) -- (0,\yr) node [below right] {$\Im j$};
  \draw[xshift=-5,blue!60!black,decoration={markings,mark=between positions 0.125 and 0.875 step 0.25 with \arrow{>}},postaction={decorate}] (-\xr,\yr/4) -- (.55\yr,\yr/4) arc (90:-90:\yr/4) (.55\yr,-\yr/4) -- (-\xr,-\yr/4);
  \poles
  \polecontours
\end{tikzpicture}}
    \caption{Complex j-plane structure for the odd spin-j bulk-to-bulk propagator. The poles give rise to the vector glueball spectrum and the cut gives rise to the Odderon.}
    \label{fig:cuts}
\end{figure}

Reverting the rescalings and coordinate transformations, we obtain the resummed scalar bulk-to-bulk propagator in the repulsive wall to be

\begin{widetext}
    \bea
    \label{SCALARlin}
 G_0(j, t, z, z')&=&-\frac{(z z')^2}{2}(3a\kappa^2zz'/2)^{\Delta_g(j)-2}\frac{\Gamma(\frac{3\Delta_g(j)+S_j-6+3\tilde t/a}{6})}{\Gamma(\Delta_g(j)-1)}\mathbb{M}(z)\mathbb{U}(z')
\eea
\end{widetext}
where we introduced the shorthand 
$$\mathbb{M}(z)=\mathbb{M}\bigg(\frac{S_j+3\Delta_g(j)-6-3\tilde t/a}{6},\Delta_g(j)-1,a\kappa^2z^2\bigg)$$ 
and similarly for $\mathbb U(z)$.
The branch cut of 
$$\Gamma\left(\frac{1}{2}+\alpha-\beta\right)=\Gamma(iy)\approx e^{-i \gamma_E y}/iy$$ 
is chosen to the left of the integration contour, along the negative real axis as in Fig.~\ref{fig:cuts}. In the large $s/|t|$ limit, the integral is dominated by the saddle point. For the Pomeron we recall that $m_5^2=0$ and we continue to carry out the Sommerfeld-Watson transform via saddle point approximation to obtain
\begin{widetext}
    \bea
    \label{SCALARrs2}
 G_2(s, t, z, z')&=&-\frac{3}{5}f^+(\lambda)\sqrt{\frac{3\mathcal{D}}{5\pi\tau}}\frac{(z z')^2}{2}e^{-\frac{a}{2}(j_\mathbb{P}(t)-2)\kappa^2(z^2+z{'}^2)}(3a\kappa^2zz'/2)
 e^{(j_{\mathbb{P}}(t)-2)\tau},
    \eea
\end{widetext}    
where $\tau=\log(\alpha' s z z')=\chi +\log zz'$ with $\chi=\log(\alpha' s)$ the rapidity, and we used that 
$\mathbb M(0,0,z)=\mathbb U(0,0,z)=1$ and 
\be
f^+(\lambda)=i+\frac{4 \sqrt{\lambda }}{3 \pi }-\frac{\pi }{4 \sqrt{\lambda }}\,.
\label{eq:EvenSig}
\ee
Note that at the saddle point the Kummer functions are $\mathbb M(0,0,z)=1,\ \mathbb U(0,0,z)=1$ and the bulk-to-bulk propagator is fully symmetric.

For the Odderon the analytically continued spin-1 and spin-0 are related by
\bea
G_1(j,t,z,z')=e^{-(j-1)(A(z)+A(z'))}\,\tilde G_0(j,t,z,z')\nonumber\\
\eea
with the 
rescaled scalar field
\bea
\tilde G_0(j, t,  z, z')&=&(zz')^{-1}G_0(j, t,  z, z').\nonumber\\
\eea
We recall that 
$$\Delta_g(j)=2+\sqrt{4+S_j}=2+m_k$$ 
and thus 
\bea
\label{oddTrs}
j_{\mathbb{O}}(t)=j_--\frac{m_k^2-1}{2\sqrt{\lambda}}+\frac{\alpha'}{2}t\,.
\eea
We expect higher order corrections in $1/\sqrt{\lambda}$ to the Odderon  intercept in~(\ref{oddTrs}), following the strong coupling analysis in the conformal limit in~\cite{Brower:2014wha}.

The multiple branches of the Odderon, which are fixed from type IIB SUGRA, are given by $m_k^2=k^2$ and $m_k^2=(k+4)^2$. The structure of \eqref{oddTrs} shows that the $k=0$ branch has either an intercept greater than one, or is super-diffusive. The branch with $k=1$ has an intercept precisely at one, and is the leading branch cut that will be picked up by the saddle-point. Remarkably, a similar Odderon trajectory was noted 
in baryon-baryon scattering in AdS with D5 branes~\cite{Avsar:2009hc}.

Performing the Sommerfeld-Watson transform via saddle-point approximation along the contour of Fig.~\ref{fig:cuts}, the bulk-to-bulk Odderon propagator is given by
\begin{widetext}
    \bea
    \label{SCALARrs1}
 G_1(s, t, z, z')&=&-\frac{3}{5}f^-(\lambda)\sqrt{\frac{3\mathcal{D}}{5\pi\tau}}\frac{(z z')}{2}e^{-\frac{a}{2}(j_\mathbb{O}(t)-1)\kappa^2(z^2+z{'}^2)}(3a\kappa^2zz'/2)e^{(j_{\mathbb{O}}(t)-1)\tau},
    \eea
\end{widetext}
where 
\be
f^-(\lambda)=i+\frac{4 \sqrt{\lambda }}{ (m_k^2-1)\pi }-\frac{(m_k^2-1)\pi}{12 \sqrt{\lambda }},
\label{eq:OddSig}
\ee
which has a pole for the leading $k=1$ branch. We note that, at large $\lambda$, \eqref{eq:EvenSig} and \eqref{eq:OddSig} coincide for the $k=2$ branch. The
conformal limit is discussed in Appendix~\ref{CFTX}


\section{\texorpdfstring{Scattering amplitudes}{Diffractive pp and ppbar scattering amplitudes}}
~\label{SEC-IV}	 
Elastic $pp$ and $p\bar p$ scattering in the Regge limit
with the exchange of Reggeons $\mathbb P, \mathbb O$ are depicted in Fig.~\ref{fig:PO} using standard Feynman graphs. 
In dual gravity, the Feynman graphs are replaced by the Witten diagrams in Fig.~\ref{fig:hPO}. The Pomeron $\mathbb P$ is
identified with a sum of massive even-spin glueballs, while the
Odderon is identified with a sum of massive odd-spin glueballs. 
The wavy-lines are the bulk-to-bulk $G_{j_\pm}(j)$ propagators defined
above. The bulk-to-boundary Dirac fermions are represented by solid lines, and follow from the chiral Kaluza-Klein modes for the bulk Dirac fermions discussed in Appendix~\ref{ap:BulkDiracRWall}.

The even and odd spin-j contributions to Fig.~\ref{fig:PO} are
\bea
{\cal A}_{pp\rightarrow pp}(s,t)
&=&\sum_{j=2,4,..} {\cal A}_{\mathbb P}(j,s,t) 
+\sum_{j=1,3,..} {\cal A}_{\mathbb O}(j,s,t)\nonumber\\
{\cal A}_{p\bar p\rightarrow p\bar p}(s,t)
&=&\sum_{j=2,4,..} {\cal A}_{\mathbb P}(j,s,t) 
-\sum_{j=1,3,..} {\cal A}_{\mathbb O}(j,s,t)\nonumber\\
\eea

\begin{widetext}
\subsection{Even spin-j}
The even spin-j contribution to diffractive $pp$ scattering in Fig.~\ref{fig:hPomeronEx} is given by 
\be
i{\cal A}_{\mathbb P}(j, s,t)=
(-i){\cal V}^{\mu\nu}_{j\Psi\Psi}(q_1,q_2, k, m_n)
\tilde{G}_{\mu\nu,\alpha\beta}(k) 
(-i){\cal V}^{\alpha\beta}_{j\Psi\Psi}(p_1,p_2, k, m_n)
\ee    
with the bulk Pomeron-nucleon vertex for the repulsive wall \cite{Mamo:2019mka}
\be
{\cal V}^{\alpha\beta}_{\mathbb{P}\Psi\Psi}(p_1,p_2, k, m_n)
=-\frac 12\sqrt{2\kappa^2} \int dz \sqrt{g}e^{-3A(z)}\,
\overline\Psi(p_2, z)\gamma^\alpha p^\beta\Psi(p_1, z)\,J_h(m_n, z))
\ee
where
\be
J_h(m_n, z)\equiv \sqrt{\frac{3}{4}a}\kappa z^3e^{-(j_\mathbb{P}(t)-2)a\kappa^2 z^2/2}
\ee
and the reduced even spin-j Reggeized graviton exchange
\be
G_{\mu\nu, \alpha\beta}(m_n, k, z, z^\prime)=J_h(m_n, z)
\tilde{G}_{\mu\nu, \alpha\beta}(m_n(j), k) J_h(m_n, z^\prime)
\ee
with the Pomeron propagator given by
\beq
\tilde{G}_{\mu\nu, \alpha\beta}(m_n k)=-\frac{3}{5}f^+(\lambda)\sqrt{\frac{3\mathcal{D}}{5\pi\tau}}e^{(j_{\mathbb{P}}(t)-2)\tau}(-i)\,P_{\mu\nu, \alpha\beta}(k)\nonumber\\
\eeq    
and
 \bea
P_{\mu\nu, \alpha\beta}(k)=
\frac 12 \bigg(P_{\mu\alpha} P_{\nu\beta}+P_{\mu\beta}P_{\nu\alpha}-\frac 23 P_{\mu\nu}P_{\alpha\beta}\bigg)(k),\qquad 
P_{\mu\alpha}(k)=-\eta_{\mu\alpha}+\frac {k_\mu k_\nu}{k^2}.
\eea   
In the Regge limit the amplitude reduces to
  \bea
    \label{ReggeAeven}
   {\cal A}_{\mathbb P}(s,t)&=&\frac{3}{5}f^+(\lambda)\sqrt{\frac{3\mathcal{D}}{5\pi\tau}}\frac{\kappa^2}{2}g_{\mathbb{P}\overline{\Psi}\Psi}^2 4s^2\left(1+\frac{t-4m_p^2}{s}\right)e^{(j_{\mathbb{P}}(t)-2)\tau}\delta_{s_1 s_2}\delta_{s_1's_2'}\nonumber\\
   &\equiv & \mathcal{N}f^+(\lambda)e^{(j_\mathbb{P}(t)-2)\tau}\mathcal{V}_\mathbb{P}(s,t),
\eea  
where
\bea  
g_{\mathbb{P}\overline{\Psi}\Psi}=\frac{729f_0m_p\kappa_N^4\kappa(n_L^2+n_R^2)}{4(3\kappa_N^2+8(j_\mathbb{P}(0)-2)\kappa^2)^5},\ \qquad\qquad\qquad
\mathcal{N}=\frac{3}{5}\sqrt{\frac{3\mathcal{D}}{5\pi\tau}}\frac{\kappa^2}{2}
\eea
and we used $a=4$ for the closed string exchange and $a=1$ for the open string fields as well as  $\overline{u}(p_2)\gamma^\mu u(p_1)=(p_1+p_2)^\mu \delta_{s_1s_2}$.

\begin{figure}[!htb]
    \begin{minipage}{.45\textwidth}
       \centering
        \begin{tikzpicture}[node distance=1cm and 1.5cm]
            \coordinate[label={[yshift=4pt]left:$\Psi(q_1;z')$}] (e1);
            \coordinate[below right=of e1] (aux1);
            \coordinate[above right=of aux1,label={[yshift=4pt]right:$\overline{\Psi}(q_2;z')$}] (e2);
            \coordinate[below=1.25cm of aux1] (aux2);
            \coordinate[below left=of aux2,label={[yshift=-4pt]left:$\Psi(p_1;z)$}] (e3);
            \coordinate[below right=of aux2,label={[yshift=-4pt]right:$\overline{\Psi}(p_2;z)$}] (e4);
            
            \draw[fermion] (e1) -- (aux1);
            \draw[fermion] (aux1) -- (e2);
            \draw[fermion] (e3) -- (aux2);
            \draw[fermion] (aux2) -- (e4);
            \draw[pomeron,double] (aux1) -- node[label={right:$G_2(s,t,z,z')$}] {} (aux2);
            \node[draw, thick, circle,fit=(e1) (e4),inner sep=.5\pgflinewidth] {};
        \end{tikzpicture}
        \center{(a)}
        \label{fig:hPomeronEx}
    \end{minipage}
    \begin{minipage}{0.45\textwidth}
    \centering
            \begin{tikzpicture}[node distance=1cm and 1.5cm]
            \coordinate[label={[yshift=4pt]left:$\Psi(q_1;z')$}] (e1);
            \coordinate[below right=of e1] (aux1);
            \coordinate[above right=of aux1,label={[yshift=4pt]right:$\overline{\Psi}(q_2;z')$}] (e2);
            \coordinate[below=1.25cm of aux1] (aux2);
            \coordinate[below left=of aux2,label={[yshift=-4pt]left:$\Psi(p_1;z)$}] (e3);
            \coordinate[below right=of aux2,label={[yshift=-4pt]right:$\overline{\Psi}(p_2;z)$}] (e4);

            \draw[fermion] (e1) -- (aux1);
            \draw[fermion] (aux1) -- (e2);
            \draw[fermion] (e3) -- (aux2);
            \draw[fermion] (aux2) -- (e4);
            \draw[Odderon] (aux1) -- node[label={right:$G_1(s,t,z,z')$}] {} (aux2);
            \node[draw, thick,circle,fit=(e1) (e4),inner sep=.5\pgflinewidth] {};
        \end{tikzpicture}
        \center{(b)}
        \label{fig:hOdderonEx}  
   \end{minipage}
   \caption{Witten diagrams for diffractive $pp$ eleastic scattering through (a) Pomeron and (b) Odderon exchange}
    \label{fig:hPO}
\end{figure}
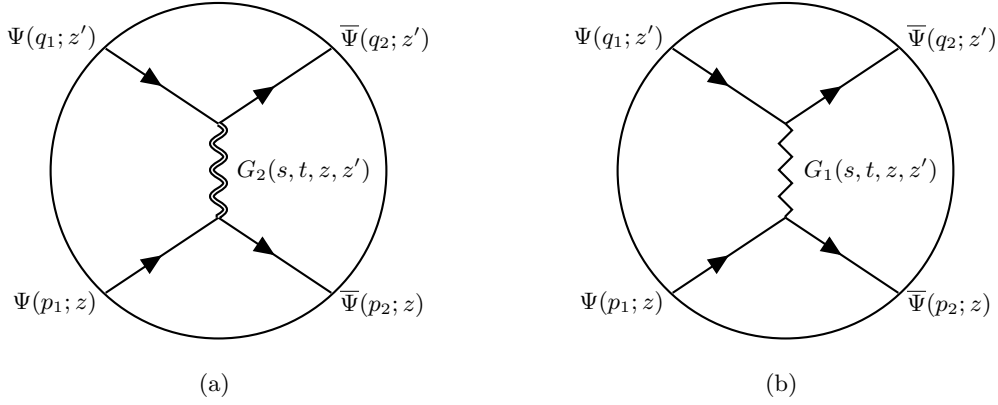

\subsection{Odd spin-j}
The odd spin-j contribution to diffractive $p\bar p$ scattering is
\be
i{\cal A}_{\mathbb O}(s,t)=\sum_{m\leq n}
(-i){\cal V}^{(n)\mu}_{j\Psi\Psi}(q_1,q_2, k, m_n)
\tilde{H}_{\mu\nu}(m_n, k) 
(-i){\cal V}^{(m)\nu}_{j\Psi\Psi}(p_1,p_2, k, m_n)
\ee
with the bulk Odderon-nucleon vertices
\bea
{\cal V}^{(1)\beta}_{\mathbb{O}\Psi\Psi}(p_1,p_2, k, m_n)
&=&+\sqrt{2\kappa^2}\frac 12 \int dz \sqrt{g}\, e^{-2A(z)}\,
\overline\Psi(p_2, z)\sigma^{\alpha\beta}\gamma^5\Psi(p_1, z)k_\alpha\,J_h(m_n, z))\\
{\cal V}^{(2)\beta}_{\mathbb{O}\Psi\Psi}(p_1,p_2, k, m_n)
&=&-\sqrt{2\kappa^2}\frac 12 \int dz \sqrt{g}\, e^{-2A(z)}\,
\overline\Psi(p_2, z)\gamma^{\beta}\gamma^5\Psi(p_1, z)\,J_h(m_n, z))
\eea
\end{widetext}
corresponding to a Pauli and Dirac coupling, where
\be
&&J_h(m_n, z)=\nonumber\\
&&\sqrt{\frac{3}{4}a}\kappa z^2e^{-(j_\mathbb{O}(t)-1)a\kappa^2 z^2/2}
\ee
and the reduced odd spin-j Reggeized spin-1 exchange
\bea
&&H_{\mu\nu}(m_n, k, z, z^\prime)=\nonumber\\
&&J_h(m_n, z)
\tilde H_{\mu\nu}(k,z,z') J_h(m_n, z^\prime),\nonumber\\
\eea   
where
\bea
&&\tilde H_{\mu\nu}(k, z, z^\prime)=\nonumber\\
&&-\frac{3}{5}f^-(\lambda)\sqrt{\frac{3\mathcal{D}}{5\pi\tau}}e^{(j_{\mathbb{O}}(t)-1)\tau}(-i)\,P_{\mu\nu}(k).\nonumber\\
\eea
The vertices arise from the sources $B_{MN}$ and $C_{MN}$ of the boundary operators in \eqref{OP1X} which are assumed to minimally couple to the chiral Dirac fermion current $\overline{\Psi}\sigma^{AB}\Psi$ in the bulk with  $\sigma^{AB}=\frac{i}{2}\left[\Gamma^A,\Gamma^B\right],\ e^M_A=e^{-A(z)}\delta^M_A,\ \Gamma^A=(\gamma^\mu,-i\gamma^5),\ \left\{\Gamma^A,\Gamma^B\right\}=2\eta^{AB}$. The corresponding $1^{--}$ fluctuations, which are tied through a topological mass term in type IIB SUGRA, are given by $C_{\mu\nu}$ and $B_{\mu z}$ \cite{Brower:2000rp}. The latter already corresponds to a spin-1 exchange, while for $C_{\mu\nu}$ a projection onto the spin-1 content is obtained via utilizing the field strength $C_{\mu\nu}=\frac{1}{2\sqrt{-\partial^2}}\epsilon_{\mu\nu\rho\sigma}F^{\rho\sigma}$. The normalization is implied by the normalized kinetic term in the closed string SUGRA action, which we omit in the following. A similar reasoning holds for $B_{\mu z}$. Note that the Pauli coupling vanishes in the forward limit, even with the normalization included, and is strongly suppressed in the Regge limit. The Dirac coupling does not and allows for the study of potential contributions of the Odderon as a spin-1 exchange in the forward region. Recall that the BKP Odderon is a Reggeized 4-vector $(\frac 12, \frac 12)$ representation 
of the complexified SO(3,1). 
For completeness, we note that in~\cite{Avsar:2009hc} the Odderon is identified with an exchange of the 
$B_{+z}$ component of the Kalb-Ramond field in $AdS_5\times S_5$. The self-dual 2-form corresponds to the $(0,1)$ representation of the complexified SO(3,1), and is found
to have non-zero coupling to the baryon vertex.

Similarly, after reducing the chiral Dirac spinors to 4D and utilizing the LSZ formula, we obtain for the Odderon
\begin{widetext}
\bea
    g^{(1)}_{\mathbb{O}\overline{\Psi}\Psi}&=&\int dz \sqrt{g}\, e^{-2A(z)}\,
\left(\psi_L(z)^2+\psi_R(z)^2\right)\,J_h(m_n, z))=\frac{7776f_0m_p\kappa_N^4\kappa(n_L^2+n_R^2)}{(9\kappa_N^2+32(j_\mathbb{O}(0)-1)\kappa^2)^4}\\
    g^{(2)}_{\mathbb{O}\overline{\Psi}\Psi}&=&\int dz \sqrt{g}\, e^{-2A(z)}\,
2\psi_L(z)\psi_R(z)\,J_h(m_n, z))=\frac{15552f_0m_p\sqrt{3\pi}\kappa\kappa_N^4n_Ln_R}{(9\kappa_N^2+32(j_\mathbb{O}(0)-1)\kappa^2)^4}
\eea
In the Regge limit the amplitude is given by
  \bea
  \label{ReggeAodd}
   {\cal A}_{\mathbb O}(s,t)&=&\frac{3}{5}f^-(\lambda)\sqrt{\frac{3\mathcal{D}}{5\pi\tau}}e^{(j_{\mathbb{O}}(t)-1)\tau}\frac{\kappa^2}{2}\Bigg(g_{\mathbb{O}\overline{\Psi}\Psi}^{(1)2}4k_\mu k_\alpha\overline{u}(p_2)\sigma^{\mu\nu}u(p_1)P_{\nu\beta}(k)\overline{u}(q_2)\sigma^{\alpha\beta}u(q_1)\nonumber\\
   &&+g_{\mathbb{O}\overline{\Psi}\Psi}^{(2)2}(p_1+p_2)^\mu P_{\mu\nu}(k)(q_1+q_2)^\nu\delta_{s_1s_2}\delta_{s_1's_2'}\nonumber\\
   &&+ig_{\mathbb{O}\overline{\Psi}\Psi}^{(1)}g_{\mathbb{O}\overline{\Psi}\Psi}^{(2)}\bigg(k_\alpha\overline{u}(q_2)\sigma^{\alpha\beta}u(q_1)P_{\beta\nu}(p_1+p_2)^\nu\delta_{s_1s_2}+k_\mu \overline{u}(p_2)\sigma^{\mu\nu}u(p_1)P_{\nu\beta}(k)(q_1+q_2)^\beta\delta_{s_1's_2'}\bigg)\Bigg)\nonumber\\
   &\equiv &\mathcal{N}f^-(\lambda)e^{(j_\mathbb{O}(t)-1)\tau}\mathcal{V}_\mathbb{O}(s,t)
\eea
where we again used $\overline{u}(p_2)\gamma^\mu u(p_1)=(p_1+p_2)^\mu \delta_{s_1s_2}$ for $s\to\infty$. In particular we obtain for the forward amplitude
  \bea
   {\cal A}_{\mathbb O}(s,0)=\mathcal{N}f^-(\lambda)\left(g_{\mathbb{O}\overline{\Psi}\Psi}^{(2)}\right)^2e^{(j_{\mathbb{O}}(t)-1)\tau}2s\left(1-\frac{4m_p^2}{2s}\right).
\eea
\end{widetext}


\subsection{Total cross sections}
For the computation of the total cross section,  we recall that the signature factors are given by 
\bea
f^+(\lambda)&=&i+\frac{4 \sqrt{\lambda }}{3 \pi }-\frac{\pi }{4 \sqrt{\lambda }}\\
f^-(\lambda)&=&i+\frac{4 \sqrt{\lambda }}{ (m_k^2-1)\pi }-\frac{(m_k^2-1)\pi}{12 \sqrt{\lambda }}.\nonumber
\eea
For the discussion of the analytical results we will drop the subleading piece in $\lambda$ and only include it in the numerical analysis. In both cases the  real part dwarfs the imaginary part at very strong coupling, but both imaginary parts are equal to 1. 

The total cross sections in $pp$ and $p\bar p$ follow from the 
optical theorem
\bea
\sigma_{\pm }(s)=
\frac {1}s{\rm Im} (\mathcal{A}_{\mathbb P}(s,0)\pm \mathcal{A}_{\mathbb O}(s,0)),
\eea
and is given by
\begin{widetext}
\bea
\sigma_\pm(s)&=&
\frac{6}{5}\sqrt{\frac{2{\cal D}}{5\pi\tau}}\frac{\kappa^2}{2}
\bigg(2
g_{\mathbb{P}\overline{\Psi}\Psi}^2\,e^{(j_{\mathbb P}(0)-1)\tau}\left(1-\frac{4m_p^2}{s}\right)
\pm 
g_{\mathbb{O}\overline{\Psi}\Psi}^{(1)2}\,e^{(j_{\mathbb O}(0)-1)\tau}\left(1-\frac{4m_p^2}{2s}\right)\bigg)\nonumber\\
&=&2\mathcal{N}e^{(j_{\mathbb P}(0)-1)\tau}\left(
g_{\mathbb{P}\overline{\Psi}\Psi}^2\left(1-\frac{4m_p^2}{s}\right)\pm 
g_{\mathbb{O}\overline{\Psi}\Psi}^{(1)2}\,e^{(j_{\mathbb O}(0)-j_{\mathbb P}(0))\tau}\left(1-\frac{4m_p^2}{2s}\right)\right),\nonumber\\
\eea
where we expanded for large $s$.
The rho-parameters  for both channels read
\bea
\label{RHOTREE}
\rho_\pm (s)=\frac{\Re\mathcal{A}_\pm(s,0)}{\Im\mathcal{A}_\pm(s,0)}=
\frac{\frac{4\sqrt\lambda}{3\pi}(1-4m_p^2/s)g_{\mathbb{P}\overline{\Psi}\Psi}^2\,e^{(j_{\mathbb P}(0)-1)\tau}\pm
\frac{4\sqrt\lambda}{(m_k^2-1)\pi}(1-4m_p^2/2s)g_{\mathbb{O}\overline{\Psi}\Psi}^2\,e^{(j_{\mathbb O}(0)-1)\tau}}
{(1-4m_p^2/s)g_{\mathbb{P}\overline{\Psi}\Psi}^2\,e^{(j_{\mathbb P}(0)-1)\tau}
\pm (1-4m_p^2/2s)g_{\mathbb{O}\overline{\Psi}\Psi}^2\,e^{(j_{\mathbb O}(0)-1)\tau}}
\rightarrow \frac{4\sqrt\lambda}{3\pi}\nonumber\\
\eea
\end{widetext}
with a constant asymptotic in the large rapidity limit.
The high energy 
scattering data suggests $\rho(s)\sim 0.1$
for $\sqrt{s}\geq 100$ GeV~\cite{Bence:2018cby} (and references therein).
Below we show that after the eikonal resummation, the strong shadowing 
caused by the exchanged Reggeons will deplete (\ref{RHOTREE}) to
zero in the large rapidity limit, in qualitative agreement with the data.



\subsection{Elastic differential cross-section}
The spin averaged squared elastic amplitude is given by 
\begin{widetext}
 \bea
\left|\mathcal{A}^\pm_E(s,t)\right|^2=\frac{s^2}{4g_5^2}& &\left[e^{2j_\mathbb{P}(t)\tau}g_{\mathbb{P}\overline{\Psi}\Psi}^4\left(1+\frac{2t-8m_p^2}{s}+\frac{21}{16}\left(\frac{t}{s}\right)^2-\frac{10m_p^2t}{s^2}\right)+4e^{2j_{\mathbb{O}}(t)\tau}\frac{\left(g^{(2)}_{\mathbb{O}\overline{\Psi}\Psi}\right)^4}{s^2}\right.\\
& &\left. \pm \frac{4}{s}(g_{\mathbb{P}\overline{\Psi}\Psi}g^{(2)}_{\mathbb{O}\overline{\Psi}\Psi})^2e^{(j_\mathbb{P}(t)+j_\mathbb{O}(t))\tau} + 64e^{2j_\mathbb{O}(t)\tau}\left(g^{(1)}_{\mathbb{O}\overline{\Psi}\Psi}\right)^4\left(\frac{t}{s}\right)^2\right]\nonumber\\
\label{eq:A2Elastic}
\eea  
\end{widetext}
where we note that the interference term is highly suppressed in the Regge limit but still leading compared to the $\left(g^{(1,2)}_{\mathbb{O}\overline{\Psi}\Psi}\right)^2$ piece. Hence we will drop $g^{(1)}_{\mathbb{O}\overline{\Psi}\Psi}$ in our numerical analysis. The elastic differential cross sections for $pp$ and $p\bar p$
follow from
\bea
\frac{d\sigma_\pm}{dt}
=\frac{1}{16\pi s^2}|\mathcal{A}_\pm (s,t)|^2
\eea
The elastic slopes are identical in the large rapidity limit
\bea
\label{BS0}
B_\pm (s,t=0)&=&\frac d{dt}\bigg({\rm ln}\,\frac{d\sigma_\pm }{dt}(s,t)\bigg)_{t=0}\nonumber\\
        &\rightarrow& 2\tau\left(\frac{d}{dt}j_{\mathbb{P}}(t)\right)_{t=0}=\alpha'\tau\nonumber\\
\eea
The  squared string length $\alpha'=1\,{\rm GeV}^{-2}$,
is comparable to the
canonical logarithmic value of the slope parameter~\cite{Bence:2018cby} (and references therein) 
\bea
B(s,t=0)=1\,{\rm GeV}^{-2}\,{\rm ln}\bigg(\frac s{1\,{\rm GeV}^2}\bigg)
\eea
for both the Pomeron and the Odderon as closed string exchanges. 

\section{Eikonal elastic scattering}
\label{SEC_EIK}
The single Pomeron and Reggeon exchanges violate unitarity
at large $\sqrt{s}$. A simple remedy is to resum the s-channel
exchanges or a process known as eikonalization, which is dominant in the Regge limit. With this in mind and 
following \cite{Brower:2007qh},  we write the holographic total cross section for the Reggeons as
\begin{widetext}
    \be
    \label{SIGMAR}
\sigma_{\mathbb R}(s)=\frac {1}s\int d^2b_\perp dzdz' 
\big(\sqrt{g(z)}\psi_{12}(j_{\mathbb R},z)\big)\big(\sqrt{g(z')}\psi_{34}(j_{\mathbb R},z')\big)\,2s\,{\rm Im}\ \chi(j_{\mathbb R},s, b_\perp,z,z')
    \ee
\end{widetext}
where $\psi_{ij}(j_{\mathbb R},z)$ are the holographic wavefunctions for the in-out states in the Regge limit,
with the metric factors $e^{-3A(z)}$ (Pomeron) and $e^{-2A(z)},\ e^{-A(z)}$ (Odderon) for each vertex implicit. 
Here $\chi(j_{\mathbb R},s,b_\perp,z,z')$ is the eikonal phase-shift for the Reggeon $\mathbb R=\mathbb P, \mathbb O$
\bea
\chi(j_{\mathbb R},s,b_\perp,z,z')=s^{j_{\pm} -1}\,
G(j_{\mathbb R},s,b_\perp,z, z')\nonumber\\
\eea
with $j_\pm=2,1$ for the Pomeron and Odderon, respectively. In the following we will use the shorthand $\chi_\mathbb{R}\equiv\chi(j_\mathbb{R},s,b_\perp,z,z')$.
More specifically, from (\ref{SCALARrs2}-\ref{SCALARrs1}) we have
\begin{widetext}
    \bea
     s^1\,G(j_\mathbb P, s, b_\perp, z, z')&=&\frac{3}{5}f^+(\lambda)\sqrt{\frac{3\mathcal{D}}{5\pi\tau}}\frac{(z z')^2}{2}e^{-\frac{a}{2}(j_\mathbb{P}(0)-2)\kappa^2(z^2+z{'}^2)}(3a\kappa^2zz'/2)
 e^{(j_{\mathbb{P}}(0)-1)\tau}\,
 e^{-\frac {b_\perp^2}{2\alpha'\tilde \tau}}\frac{4\pi}{\alpha'\tilde\tau}
 \nonumber\\
s^0\, G(j_\mathbb O, s, b_\perp, z, z')&=&\frac{3}{5}f^-(\lambda)\sqrt{\frac{3\mathcal{D}}{5\pi\tau}}f^-(\lambda)\frac{(z z')}{2}e^{-\frac{a}{2}(j_\mathbb{O}(0)-1)\kappa^2(z^2+z{'}^2)}(3a\kappa^2zz'/2)e^{(j_{\mathbb{O}}(0)-1)\tau}\,
 e^{-\frac {b_\perp^2}{2\alpha'\tilde \tau}}\frac{4\pi}{\alpha'\tilde\tau}\nonumber\\
    \eea
\end{widetext} 
with 
$$\tilde\tau=\tau\bigg(1-\frac 1{5\tau}\frac {z^2+z'^2}{R^2}\bigg)$$

The eikonalized elastic amplitudes for $pp$ and $p\bar p$ scattering in impact parameter space 
are
\bea
{\cal A}_{pp}(s, b_\perp, z, z')&=&-2is\,\bigg(e^{i(\chi_{\mathbb P}+\chi_{\mathbb O})}-1\bigg)
\nonumber\\
{\cal A}_{p\bar p}(s, b_\perp, z, z')&=&-2is\,\bigg(e^{i(\chi_{\mathbb P}-\chi_{\mathbb O})}-1\bigg)
\nonumber\\
\eea
by analogy with the scattering amplitude in quantum mechanics, with 
$\frac{1}{2}\chi_\pm\equiv\frac{1}{2}(\chi_\mathbb{P}\pm\chi_\mathbb{O})$ playing the role of the phase shifts.
Note that when reverted to momentum space, the eikonalized elastic amplitudes are
\begin{widetext}
\bea
\label{eq:eikonAmp}
{\cal A}_{pp}(s,t)&=&-2is\,\int d^2b_\perp\,e^{-iq_\perp\cdot b_\perp}\,
\bigg(e^{i(\chi_{\mathbb P}+\chi_{\mathbb O})}-1\bigg)
\nonumber\\
{\cal A}_{p\bar p}(s,t)&=&-2is\,\int d^2b_\perp\,e^{-iq_\perp\cdot b_\perp}\,\bigg(e^{i(\chi_{\mathbb P}-\chi_{\mathbb O})}-1\bigg)
\eea

To evaluate \eqref{eq:eikonAmp} we write 
\bea
\chi_{\mathbb{R}}(s,b_\perp,z,z')&=&a_{\mathbb{R}}e^{-\frac{b_\perp^2}{2\alpha'\tilde{\tau}}}\nonumber\\
a_\mathbb{P}&=&\tilde{\mathcal{N}}f^+(\lambda)3a\kappa^2\frac{(zz')^3}{2}e^{-\frac{a}{2}(j_\mathbb{P}(0)-2)\kappa^2(z^2+z{'}^2)}
 e^{(j_{\mathbb{P}}(0)-1)\tau}s^{-j_+}\mathcal{V}_\mathbb{P}(s,t,z,z')\nonumber\\
 a_\mathbb{O}&=&\tilde{\mathcal{N}}f^-(\lambda)3a\kappa^2\frac{(z z')^2}{2}e^{-\frac{a}{2}(j_\mathbb{O}(0)-1)\kappa^2(z^2+z{'}^2)}e^{(j_{\mathbb{O}}(0)-1)\tau}s^{-j_-}\mathcal{V}_\mathbb{O}(s,t,z,z')\nonumber\\
 \tilde{\mathcal{N}}&=&\frac{6}{5\alpha'}\sqrt{\frac{3\pi\mathcal{D}}{5\tau^3}}
\eea
with $\mathbb {R=P,O}$ and $\mathcal{V}_\mathbb{P}(s,t,z,z'),\ \mathcal{V}_\mathbb{O}(s,t,z,z')$ the unaveraged versions of the vertex and kinematical factors defined in \eqref{ReggeAeven} and \eqref{ReggeAodd}, respectively. 
Defining $a_\pm=a_{\mathbb{P}}\pm a_{\mathbb{O}}$, we continue to carry out the angular integration to get
\bea
\label{APMZ}
{\cal A}_\pm (s, t, z, z')
=-4is\pi\int b db J_0(qb)\sum_{n=1}^\infty\frac{\left(ia_\pm\right)^n}{n!}e^{-\frac{n b^2}{2\alpha'\tilde{\tau}}}
=-4i\pi s\alpha'\tilde\tau\, \sum_{n=1}^\infty\frac{\left(ia_\pm\right)^n}{n\,n!}\,e^{-\frac{\alpha'}{2n}q^2\tilde{\tau}}
\eea
\end{widetext}
The exchange of sum and integration follows from the absolute convergence of the series
and the integrability of the ensuing function.  For numerical evaluations, we will mostly use
(\ref{eq:eikonAmp}), for faster numerical convergence.

\subsection{Eikonalized cross sections}
In the forward limit, the remaining sum in (\ref{APMZ})  can be carried out analytically with the result
\bea
\sum_n\frac{\left(ia_\pm\right)^n}{n\cdot n!}&=&-\left({\rm ln}(-ia_\pm)+\gamma_E+\Gamma(0,-ia_\pm)\right),\nonumber\\
\eea
where $\Gamma(a,b)$ is the incomplete Gamma function.
For fixed $z,z'$, the forward scattering amplitudes are then 
\bea
\label{APM}
{\cal A}_\pm(s, 0, z, z')=4i\pi\alpha's\tilde\tau\bigg({\rm ln}(-ia_\pm)+\gamma_E+{\cal O}(e^{ia_\pm})\bigg)\nonumber\\
\eea
where the incomplete Gamma function is seen to be suppressed 
exponentially, in the large rapidity limit. 

In terms of (\ref{APM}),
the total cross sections (\ref{SIGMAR}) are
\bea
    \label{SIGMARX}
\sigma_{\pm}(s)=\left<\frac 1s\,{\rm Im}\bigg(
{\cal A}_{\mathbb P}(s,0, z, z')\pm {\cal A}_{\mathbb O}(s,0, z, z')\bigg)
\right>\nonumber\\
\eea
where the averaging is over the in-out states using 
\bea
\label{WEIGHT}
\int dzdz' 
\big(\sqrt{g(z)}\psi_{12}(j_{\mathbb {R}},z)\big)\big(\sqrt{g(z')}\psi_{34}(j_{\mathbb {R}},z')\big)\nonumber\\
\eea
for each of the Reggeon exchanged $\mathbb R=\mathbb {P,O}$. The averaging procedure in \eqref{SIGMARX} amounts to replacing $a_\pm$ by
\begin{widetext}
\bea
\left<a_\pm\right>=\tilde{\mathcal{N}}\left(e^{(j_\mathbb{P}(0)-1)\tau}f^+(\lambda)s^{-j_+}\mathcal{V}_\mathbb{P}(s,t)\pm e^{(j_\mathbb{O}(0)-1)\tau}f^-(\lambda)s^{-j_-}\mathcal{V}_\mathbb{O}(s,t)\right).\nonumber\\
\eea
Hence, the total cross sections are
\bea
\sigma_\pm (s)=4\pi\alpha'\tilde \tau \,{\rm Re}\,\bigg(
{\rm ln}(-i\left<a_\pm\right>)+\gamma_E+{\cal O}(e^{i\left<a_\pm\right>})\bigg),\nonumber\\
\eea
or, more explicitly,
    \bea
\sigma_\pm (s)= 4\pi\alpha'\tau\bigg((j_{\mathbb P}(0)-1)\tau-\frac{3}{2}{\rm ln}\tau
+{\rm ln}\big|h_+\pm h_-\,e^{(j_{\mathbb O}(0)-j_{\mathbb P}(0))\tau}\big|+\gamma_E+{\cal O}
\bigg)
    \eea
    \end{widetext}
    with
    \bea
    h_\pm=s^{-j_\pm}\mathcal{V}_\mathbb{R}(s,0)\,f^\pm(\lambda)\equiv \bar h_\pm\, f^\pm (\lambda)\nonumber\\
    \eea
which is seen to asymptote the Froissart bound in the large rapidity limit
\bea
\label{FROISSART}
\sigma_\pm (s)\rightarrow 4\pi\alpha'\,(j_{\mathbb P}(0)-1)\,\tau^2 
\eea
in conformity with unitarity.
The bound is fixed by the Pomeron intercept, which is larger than the highest Odderon intercept for $k=1$.

The difference between the $pp$ and $p\bar p$ cross sections, is exponentially small  at large rapidities 
\bea
\sigma_+-\sigma_-&=&
4\pi\alpha'\tau\,
{\rm ln}\bigg|
\frac{1+\frac{h_-}{h_+} e^{(j_{\mathbb O}-j_{\mathbb P})\tau}}
{1-\frac{h_-}{h_+} e^{(j_{\mathbb O}-j_{\mathbb P})\tau}}\bigg|\nonumber\\
&\rightarrow& 
8\pi\alpha'\tau\,\frac{1+\frac{16\lambda}{3(m_k^2-1)\pi^2}}{1+\frac{16\lambda}{9\pi}}\,
\,\frac{\bar h_-}{\bar h_+}\,e^{(j_{\mathbb O}(0)-j_{\mathbb P}(0))\tau}\nonumber\\
\eea
The $pp$ cross section is smaller than the $p\bar p$ cross section for the Odderon branch $m_k^2=k^2\rightarrow 0$, which yields the Odderon trajectory with intercept $j_{\mathbb O}(0)=1+\frac{1}{2\sqrt{\lambda}}$, which is, however, not picked up by the contour in Fig.~\ref{fig:cuts}.  This situation is reversed for the higher $k>1$ contributions and also the branch $m_k^2=(4+k)^2$, which are Odderons with intercepts below 1. 




\subsection{rho and B parameters}
The corresponding rho-parameters following from the eikonal amplitudes, are now given by
\bea
\rho_\pm(s) =\frac{\rm Re\,{\cal A}_\pm(s,0)}{\rm Im\,{\cal A}_\pm(s,0)}=
\frac{-\frac 12 {\rm Im}\,{\rm ln}(-i\left<a_\pm\right>/{+i\left<a_\pm\right>^*})}
{{\rm Re}\,{\rm ln}|\left<a_\pm\right>| +\gamma_E+{\cal O}}\nonumber\\
\label{rhoEikon}
\eea
In particular, at large rapidity
\bea
\label{RHOPM}
&&\rho_+(s)\rightarrow \frac {\frac \pi 2(1+{\cal O}(\frac 1{\sqrt\lambda}))}{(j_{\mathbb P}(0)-1)\,\tau}\nonumber\\
&&\rho_-(s)\rightarrow \frac {{\cal O}(\frac 1{\sqrt\lambda})}{(j_{\mathbb P}(0)-1)\,\tau}
\eea
which asymptote 0.
This is  seen to follow from the strong shadowing brought about by the eikonal resummation.

To proceed numerically, we fix the strong 't Hooft coupling by setting the soft Pomeron intercept to the Luscher contribution from the Nambu-Goto string
time-like~\cite{Liu:2018gae} 
\be
2-\frac 3{2\sqrt\lambda}=1+\frac 16
\ee
which is close to
the soft phenomenological Donnachie-Landshoff Pomeron intercept~\cite{Donnachie:1984xq}. Remarkably $\frac 16$ is precisely the
entanglement entropy of free bosons (string bits) with a fixed (open) boundary in 1+3-dimensions. Further, due to the intercept of the Odderon being below 1, as enforced by the signature factor $f^-(\lambda)$, the dependence on the Odderon coupling $g_{\mathbb{O}\overline{\Psi}\Psi}^{(2)}$ is very subtle. We thus refrain from a fit of $g_{\mathbb{O}\overline{\Psi}\Psi}^{(2)}$ to the data and instead use different input values to obtain an upper bound on the coupling, which will later serve as input for the differential cross section.

\begin{figure}[H]
    \centering
    \includegraphics[height=5cm,width=8cm]{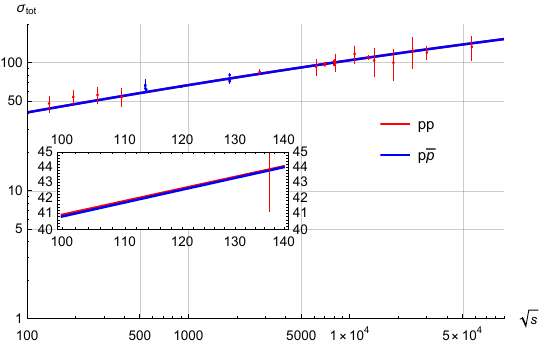}
    \caption{Total cross sections for $pp$ and $p\bar p$ scattering, with the parameters given in~Tab.~\ref{tab:bestFit}.}
    \label{fig:sigTotFit}
\end{figure}
\begin{figure*}
\subfloat[\label{fig:rhoFit}]{%
\includegraphics[height=5cm,width=.45\linewidth]{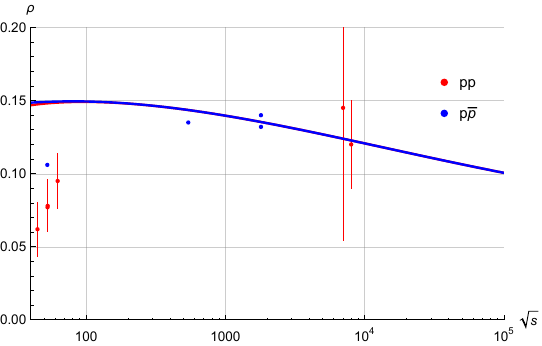}%
}\hfill
\subfloat[\label{fig:slopeFit}]{%
  \includegraphics[height=5cm,width=.45\linewidth]{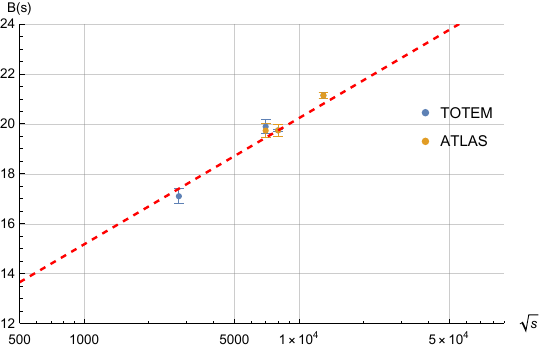}%
}
\caption{Rho (a) and slope (b) parameters versus $\sqrt{s}$.}
\label{fig:slopeFit1}
\end{figure*}

In Fig.~\ref{fig:sigTotFit} we show the total cross section for $pp$ and $p\bar p$ for a global fit to the $pp,\ p\bar p$ total cross sections, rho parameter and $B(s)$ data with $\sqrt{s}\geq 1\text{ TeV}$ \cite{TOTEM:2013lle,TOTEM:2016lxj,TOTEM:2017asr, TOTEM:2017sdy, ATLAS:2014vxr, ATLAS:2016ikn, ATLAS:2022mgx,ParticleDataGroup:2022pth}.
The parameters are
listed in Tab.~\ref{tab:bestFit}. The fit results confirm the very subtle dependence on the Odderon coupling. For the plots we fixed $g_{\mathbb{O}\overline{\Psi}\Psi}^{(2)}=15$, which minimized the mean-squares error and simultaneously showed the best numerical behaviour in the rho parameter, which is only poorly constrained by the data in the considered energy regime. In addition to the parameters $\alpha',\ g_{\mathbb{P}\overline{\Psi}\Psi}$ we introduced an overall scale factor $\cal{N_\sigma}$ for the total cross section and $\cal{N_\rho}$ for the rho parameter. A more meaningful fit will follow from the differential cross sections below.
\begin{table}[H]
    \centering
    \begin{tabular}{l|cccc}
        $g_{\mathbb{O}\overline{\Psi}\Psi}^{(2)}$ & $\alpha'$ (${\rm GeV}^{-2})$ & $g_{\mathbb{P}\overline{\Psi}\Psi}$ &$\mathcal{N}_\sigma$ &  $\mathcal{N_\rho}$\\
        \hline
        0 & 1.098(2) & 2.1856(40) & $4.6\cdot 10^{-3}(07)$ & 0.787(190)\\
        15 & 1.098(2) & 2.1856(40) & $4.6\cdot 10^{-3}(07)$ & 0.787(190)\\
        25 & 1.098(2) & 2.1857(40) & $4.6\cdot 10^{-3}(07)$ & 0.787(190)
    \end{tabular}
    \caption{Best-fit parameters for forward quantitites in $pp$ and $p\bar p$ scattering for different input values of $g_{\mathbb{O}\overline{\Psi}\Psi}^{(2)}$. See text.}
    \label{tab:bestFit}
\end{table}
In Fig.~\ref{fig:slopeFit} we show the empirical fit for the elastic slope parameter in \eqref{BS0}.
Note that the slope parameter can be recast as a measure of the rms radius
\bea
B_\pm (s)\rightarrow \frac 12\langle b^2_\perp\rangle_\pm (s)
\eea
where the averaging is understood  using the T-matrix (\ref{DIFF3}) below.
Recall that for a Gaussian T-matrix
\bea
e^{-\frac 12 \frac{b^2}{\langle b^2_\perp\rangle_\pm (s)}}
\rightarrow e^{B_\pm (s)t}
\eea
We have also introduced an overall scale ${\cal N}_\rho$ in the numerical analysis of the 
$\rho$ parameter, as listed in  Tab.~\ref{tab:bestFit}.
In \ref{fig:rhoFit} we show the fit results for the $\rho$ parameter in \eqref{rhoEikon}. 

\subsection{Elastic differential cross section}
Away from the forward limit, the spin averaged elastic differential cross sections in the eikonal approximation,  are
\bea
\label{DIFF1}
\frac {d\sigma_\pm (s,t)}{dt}=
\frac 1{16\pi s^2} 
\big<\big|{\cal A}_\pm(s, t, z, z')\big|^2\big>
\eea
with the Reggeon amplitudes for fixed impact parameter and $z,z'$ given in (\ref{APMZ}), and the averaging carried using the weight (\ref{WEIGHT}). 
More specifically,
\bea
\frac {d\sigma_\pm (s,t)}{dt}&=&
\frac{(2\pi\alpha'\tilde\tau)^2}{16\pi s^2}\,\sum_{m,n=1}^\infty
\frac{\langle g^m_\pm\,a_\pm^{*n}\rangle}{m\,n\,m!\,n!}
\,e^{\frac{\alpha' t\tilde\tau}2(\frac 1m+\frac 1n)}\nonumber\\
&\rightarrow&
\frac{(2\pi\alpha'\tilde\tau)^2}{16\pi s^2}
\, \langle |a_\pm|^2\rangle \, e^{\alpha't\tau}
\eea
with the rightmost result following from the large rapidity limit. 
Note that the  spin-tracing is trivial in the eikonal limit.
It follows that the slope parameter (\ref{BS0}) is unchanged after the eikonal resummation, at asymptotic rapidities. 



To understand the diffractive nature of the elastic differential cross section, we can recast
(\ref{DIFF1}) in the form
\begin{widetext}
\bea
\label{DIFF2}
\frac {d\sigma_\pm (s,t)}{dt}=
\frac 1{4\pi} 
\bigg<\bigg|\int d^2b_\perp\,e^{-iq_\perp\cdot b_\perp}
\big(1-e^{i(\chi_{\mathbb P}\pm \chi_{\mathbb O})}\big)(s, b_\perp, z, z')\bigg|^2\bigg>
\eea
\label{eq:dsigEikon}
\end{widetext}
with $t=-q_\perp^2$. For large $\sqrt s$, the Froissart bound (\ref{FROISSART}) is reached
in the eikonal approximation. In this limit, we may approximate the Fourier inverse of the T-matrix in (\ref{DIFF2})
by a black disc 
\bea
\label{DIFF3}
\langle{\rm Re}{\cal T}_\pm (s, b_\perp)\rangle&=&
\langle{\rm Re}(1-e^{i(\chi_{\mathbb P}\pm \chi_{\mathbb O})}\big)(s, b_\perp)\rangle\nonumber\\
&\rightarrow &\theta (b(s)-|b_\perp|)
\eea
with a growing radius with rapidity
\bea
\label{BMAX}
b(s)=\sqrt{2\alpha'(j_{\mathbb P}(0)-1)}\tau
\eea
for both $pp$ and $p\bar p$.

\begin{figure*}
\subfloat[\label{fig:tMatrix}]{%
  \includegraphics[height=5cm,width=.45\linewidth]{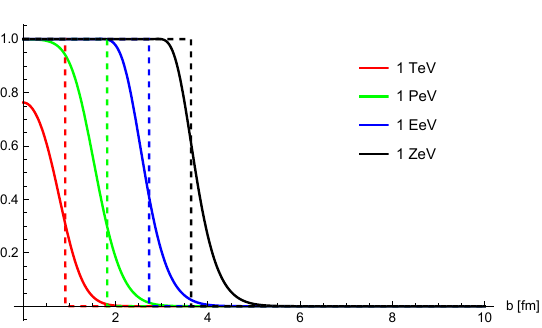}%
}\hfill
\subfloat[\label{fig:tMatrixSat}]{%
  \includegraphics[height=5cm,width=.45\linewidth]{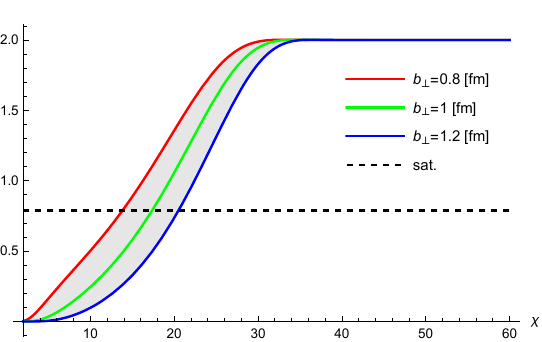}%
}
\caption{ a: the real part of the $pp$ T-matrix approaches a step function at large rapidities and large $\lambda$;
b: the differential $pp$ cross section in \eqref{eq:dSigmad2b} crosses the saturation dashed-line at large rapidities and fixed $b_\perp$ for $pp$.}
\end{figure*}
In Fig.~\ref{fig:tMatrix} we show the behavior of T-matrix versus $b_\perp$ for fixed $\sqrt{s}$, with vertices set to 1. The black-disc limit is reached for $\sqrt{s}=1\,{\rm PeV}$. The parameters used are those listed in Tab.~\ref{tab:bestFitdSigma} for $g_{\mathbb{O}\overline{\Psi}\Psi}=15$. They follow from 
a global fit to the empirical differential cross sections from
TOTEM \cite{TOTEM:2011vxg,TOTEM:2015oop, TOTEM:2018psk, TOTEM:2018hki, TOTEM:2021imi} and \D0 \cite{D0:2012erd}.
Note that the  ratio $\tau/\sqrt{\lambda}$, which we argued to be large to carry out the integrals via saddle point approximation above, is between $4.2$ and $5.3$ for the data sets used in the fits.

In Fig.~\ref{fig:tMatrixSat} we
show the behavior of the differential cross section  
\bea
 \frac{d\sigma_+}{d^2b_\perp}=
 2\langle{\rm Re}{\cal T}_+(s, b_\perp)\rangle
 \label{eq:dSigmad2b}
 \eea
versus rapidity $\chi={\rm ln}s$, for $b_\perp=0.8,1,1.2\,{\rm fm}$.
The dashed line refers to the saturation line set by the condition
\bea
 \frac{d\sigma_+}{d^2b_\perp}\bigg|_{S}=
 2(1-e^{-\frac 12})=0.79
 \eea
For $b_\perp=0.8,1,1.2\,{\rm fm}$ the crossing takes place in the rapidity range $\chi_S=14-20$, in agreement with a recent estimate using the standard  Nambu-Goto string~\cite{Liu:2023zno}
(see their Fig.~2).

\begin{figure*}
\subfloat[\label{fig:dsigppFit}]{%
\includegraphics[height=5cm,width=.45\linewidth]{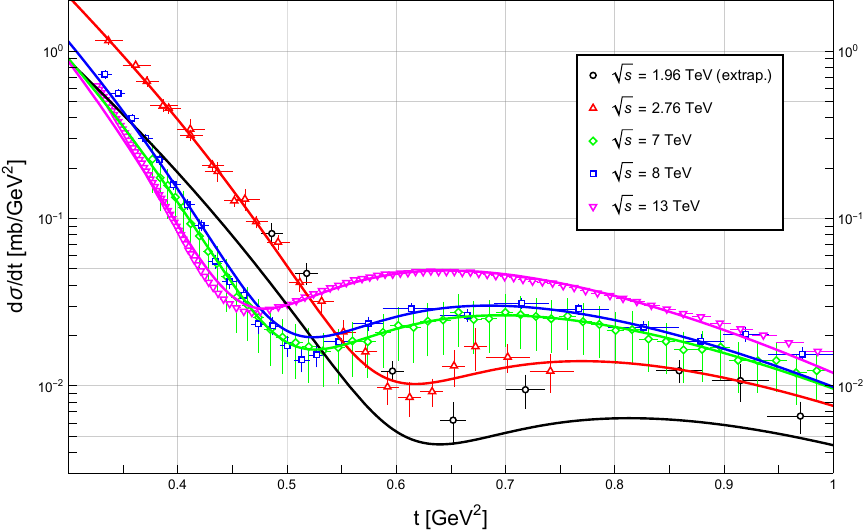}%
}\hfill
\subfloat[\label{fig:dsigppbarFit}]{%
\includegraphics[height=5cm,width=.45\linewidth]{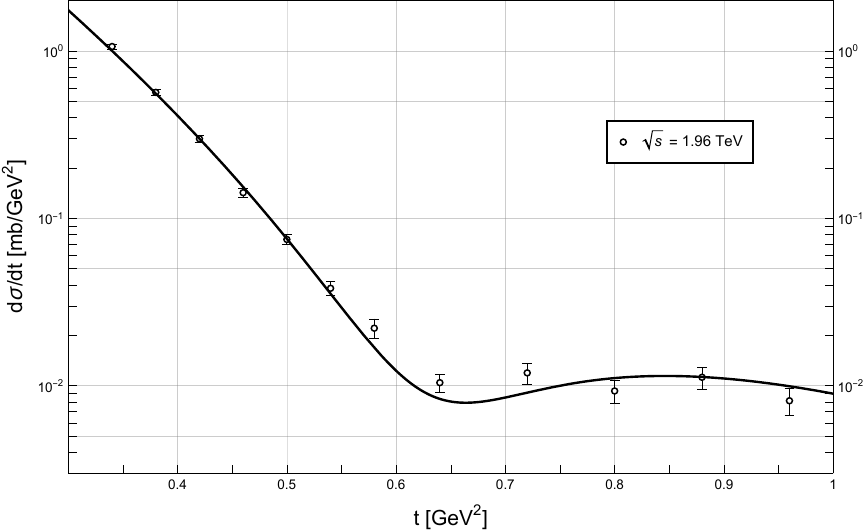}%
}
\caption{ a: Results for the differential $pp$ cross section (\ref{DIFF2}), together with a linear extrapolation of the results to $\sqrt{s}=1.96$ TeV, with a comparison to the corresponding TOTEM data and extrapolation \cite{TOTEM:2011vxg,TOTEM:2015oop, TOTEM:2018psk, TOTEM:2018hki, TOTEM:2021imi};
b: Results for the differential $p\bar{p}$ cross section (\ref{DIFF2}), compared  to the data from \D0 \cite{D0:2012erd}.}
\end{figure*}

Inserting (\ref{DIFF3}) into (\ref{DIFF2}) yields asymptotically
\bea
\label{DIFF4}
\frac {d\sigma_\pm (s,t)}{dt}\rightarrow \pi b^2(s)\,\frac{J_1^2(\sqrt{|t|}\,b(s))}{|t|}
\eea
with the expected diffraction oscillations. Note that in the black-disc limit, the first minimum in (\ref{DIFF4}) corresponds to
\bea
t_{\rm min}(s)=\frac{14.67}{b^2(s)}
\eea
The diffractive minimum shifts down with increasing $\sqrt{s}$, a pattern that is consistent with the reported  measurements in Fig.~\ref{fig:dsigppFit}. We can readily check that (\ref{DIFF4}) recovers the Froissart bound. Indeed, the total cross sections $\sigma_\pm (s)$ are tied to the elastic differential cross sections  by
\bea
\label{FROISSARTX}
\sigma_\pm(s) =\bigg(\frac{16\pi}{1+\rho_\pm^2(s)}
\bigg(\frac {d\sigma_\pm (s,t)}{dt}\bigg)\bigg)^{\frac 12}_{t=0}\rightarrow 2\pi b^2(s)\nonumber\\
\eea
where the rightmost result follows asymptotically, since the rho-parameters in (\ref{RHOPM}) vanish in this limit. Recall that in the  the black-disc limit, the elastic and inelastic cross sections are equal to $\pi b^2(s)$.  (\ref{FROISSARTX}) is in agreement with (\ref{FROISSART}). 

\begin{table}[H]
\centering
\begin{tabular}{c|ccc}
    $\sqrt{s}$   &  $\alpha'\ [\rm{GeV}^{-2}]$ & $g_{\mathbb{P}\overline{\Psi}\Psi}$ & $\mathcal{N}_{d\sigma}$\\
    \hline
    1.96 TeV   &  0.640(21)  & 1.071(15) &  0.003  \\ 
    2.76 TeV   &  0.715(27)  & 1.009(3)  &  0.007  \\ 
     7 TeV     &  0.607(5)   & 1.089(3)  &  0.002  \\
     8 TeV     &  0.626(15)  & 1.046(9)  &  0.003  \\
     13 TeV    &  0.587(5)   & 1.0782(3) &  0.002  \\
\end{tabular}
\caption{Best fit parameters for the  differential cross section data from TOTEM \cite{TOTEM:2011vxg,TOTEM:2015oop, TOTEM:2018psk, TOTEM:2018hki, TOTEM:2021imi} and \D0 \cite{D0:2012erd} with a fixed Odderon coupling of $g_{\mathbb{O}\overline{\Psi}\Psi}^{(2)}=15$. The standard error on $\mathcal{N}_{d\sigma}$ is negligible.}
\label{tab:bestFitdSigma}    
\end{table}

\begin{figure}
    \centering
    \includegraphics[height=5cm,width=8cm]{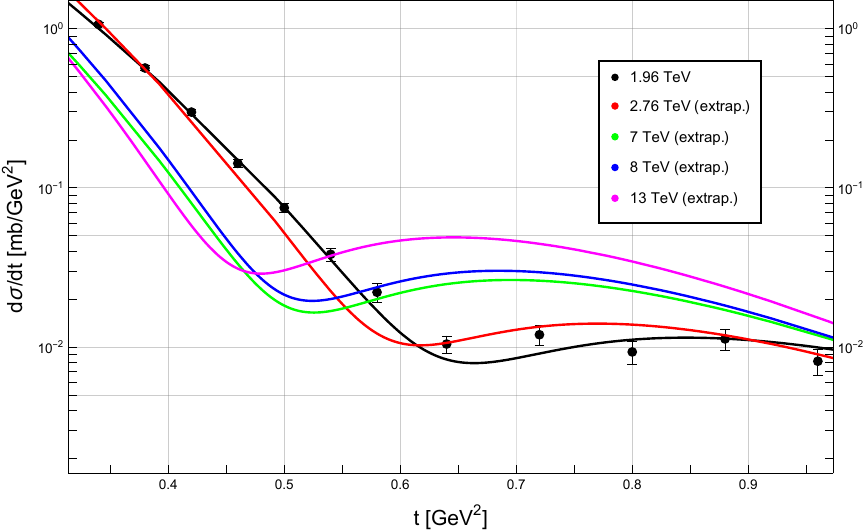}
    \caption{Extrapolated differential $p\bar{p}$ cross section}
    \label{fig:dsigppbarExtrap}
\end{figure}

In Fig. \ref{fig:dsigppFit} we show the fit results for the holographic eikonalized elastic differential $pp$ cross section \eqref{DIFF2} at center of mass energies of $\sqrt{s}=2.76,\ 7,\ 8,\ 13$ TeV \cite{TOTEM:2011vxg,TOTEM:2015oop, TOTEM:2018psk, TOTEM:2018hki, TOTEM:2021imi},  as well as  a linear extrapolation of the fit parameters from Tab.~\ref{tab:bestFitdSigma} with weighted errors to $\sqrt{s}=1.96$ TeV. The extrapolation is in qualitative agreement with Fig.~\ref{fig:TODO}. The diffractive tail is very well reproduced for the scattering data with $\sqrt{s}\geq 7$ TeV. For the data sets above $\sqrt{s}=7$ TeV, the model parameters seem to be converging to common values. The bump-dip region is not well pronounced, and with relatively large errors in the scattering data sets from TOTEM at $\sqrt{s}=2.76$ TeV. We performed the same fit with different input values for $g_{\mathbb{O}\overline{\Psi}\Psi}^{(2)}$ and found that, due to the low intercept, the dependence on this coupling is even less pronounced as for the forward quantities in Tab.~\ref{tab:bestFit}.

Our result for $p\bar p$ at $\sqrt{s}=1.96$ TeV are shown in Fig.~\ref{fig:dsigppbarFit}, and compared 
to the results reported by the \D0 collaboration. We note that the 
diffractive peak is almost absent in this channel and at this 
center of mass energy.  
In Fig.~\ref{fig:dsigppbarExtrap} we show the extrapolations of the $p\bar{p}$ differential cross section to $\sqrt{s}=2.76,\ 7,\ 8,\ 13$ TeV, with the diffractive pattern still visible. Data at these higher energies will be welcome.

The agreement of our holographic results for $pp$ and $p\bar p$ elastic cross sections at large center of mass energies with the reported TOTEM data,
does not support the contributions of both a Pomeron and Odderon. Which is, however, mostly due to not being able to fix the Odderon intercept at 1.





\section{Conclusions}					
\label{CONCLUSION}
\label{SEC-VI}
At large center of mass energies, the elastic $pp$  and $p\bar p$ amplitudes receive contributions from C-even  Pomeron and C-odd Odderon exchanges. In dual gravity, these exchanges follow from the Reggeization of the graviton and the Kalb-Ramond bulk fields~\cite{Brower:2008cy}.

We have used a bottom-up approach gravity dual construction with a repulsive wall, with bulk gravitons, vector and Dirac fields. The vector
fields, arising from the form fields $B_2$ and $C_2$, are allowed to couple through a Pauli coupling as expected for antisymmetric fields. We recall that the BKP Odderon in QCD,  is vector-like, with a non-vanishing forward coupling, as supported by our computation.

Using the leading Witten diagrams in dual gravity, we have explicitly derived the diffractive scattering amplitudes for both $pp$ and $p\bar p$ scattering in the Regge limit. The graviton-nucleon 
coupling is tensorial (one coupling), while the Odderon-nucleon coupling is vectorial (two couplings). The resulting amplitudes were eikonalized by resumming the leading Pomeron and Odderon contributions.

The model is characterized by the standard dual gravity parameters $\alpha', \lambda, g_5$ and
$\kappa$ the size of the confining and repulsive wall. We further introduced $\kappa_N$ to account for the different mass scale involved for the nucleon trajectory. These parameters are in principle fixed by 
the nucleon mass and even $2^{++}$ and odd  $1^{--}$ glueball trajectories, but we have elected to trade them for the Pomeron and Odderon couplings and fixed them by
the scattering data. 

The eikonalized holographic results for the forward $pp$ and $p\bar p$ slope and rho parameters are in relatively good agreement with the world data, including the recently reported data from TOTEM. 
Remarkably, the differential cross sections for elastic $pp$ scattering in the range $\sqrt{-t}< 1 {\rm GeV}$, are well reproduced for all reported center of mass energies including the most recent TOTEM  result at $\sqrt{s}=13\,{\rm TeV}$, with a very weak dependence on the Odderon coupling. The agreement with the data seems to be better with larger $\sqrt s$. The first diffractive oscillations (dip-bump) are reproduced.
In our model these diffractive patterns require only a strong Pomeron exchange.

To the order in $1/\sqrt\lambda$ considered and due to the pole in $f^-(\lambda)$, we are unable to fix the Odderon intercept at 1, which would in principle allow for a stronger dependence of the amplitude on the Odderon coupling. Hence, at high energies, its contribution is obscured by the Pomeron. The spurious pole at an intercept of 1 appears to be independent of the chosen background, as shown in~\eqref{eq:EvenSig},\eqref{eq:OddSig} and (\ref{eq:ConfEvenSig}-\ref{eq:ConfOddSig}). This may be
removed by retaining further higher order corrections in $1/\sqrt\lambda$, using the strong coupling analysis in~\cite{Brower:2014wha}.

In contrast, our eikonalized holographic 
differential cross section for $p\bar p$
appears to reproduce the empirical results 
quite well for $\sqrt{-t}< 1\,{\rm GeV}$ and already at the center of mass energy of $\sqrt{s}=1.96\,{\rm TeV}$. Our holographic analysis in the $p\bar p$ channel exhibits the same diffractive pattern as in the $pp$ channel. This is chiefly due to the absence of a cancellation between the Pomeron and Odderon exchange contributions with an intercept below 1. Therefore, we conclude that an underlying holographic Odderon exchange is not present in the currently reported TOTEM data.

In addition, the empirical TOTEM data for diffractive $pp$ scattering indicate that
the proton saturates for rapidities in the range $14<\chi_S< 20$, in overall agreement with a recent estimate using the Nambu-Goto string in 4-dimensions~\cite{Liu:2023zno}.
The lower bound translates to parton-x saturation for $x_S>10^{-6}$ in deep inelastic scattering, a challenging range for the upcoming electron-ion colliders.

Finally, the questions we raised in the introduction can now be answered in the context of dual gravity: 1/ the persistent diffractive structure at low-t in the $pp$ data is due to the exchange of mostly eikonalized Pomerons with strong shadowing; 2/ The absence of secondary structure at large-t is due to strong shadowing; 3/ The linear rise of the forward slope measures is a measure of the rise of the Pomeron slope which is intimately related to the rise of the $2^{++}$ glueball Regge trajectory.

\vskip 1cm
\noindent{\bf Acknowledgements:}
\vskip 0.5cm
\noindent This work  is supported by the Office of Science, U.S. Department of Energy under Contract No. DE-FG-88ER40388. It is also supported in part within the framework of the Quark-Gluon Tomography (QGT) Topical Collaboration, under contract no. DE-SC0023646.
F.~H. has been supported by the Austrian Science Fund FWF, project no. P 33655-N and the FWF doctoral program Particles \& Interactions, project no. W1252-N27. K.M. is supported by U.S. DOE Grant No. DE-FG02-04ER41302.

\appendix
\section{Bulk-to-bulk propagator in the soft wall model}
\label{ap:b2bSW}
For comparison with the repulsive wall results used earlier, we will
present a detailed derivation for the bulk-to-bulk scalar propagator 
using the soft wall model. While
many of the repulsive wall features are recovered, Gribov diffusion in the non-conformal limit is not.
To see this, consider
the soft wall model with the fixed dilaton profile $\phi(z)=(2\kappa z)^2$, the Reggeized scalar propagator associated to the Sturm-Liouville problem $L_zy(z)=\frac {\delta(z-z')}{w(z)}$
with
\bea
L_z=\frac 1{w(z)}\,d_z(w(x)\,p_0(z)\,d_z)+p_2(z),
\eea
and
\bea
w(z)&=&\sqrt{g}e^{-\phi(z)}\nonumber\\
p_0(z)&=&-g^{zz}(z)\nonumber\\
p_2(z)&=&S_j-tz^2,
\eea
with  $S_j=m_5^2R^2+m_j^2R^2$, is solution to 
\begin{widetext}
    \bea
    \label{KDIFF1}
    \bigg(-z^3e^{4\kappa^2z^2}\partial_z\bigg(\frac 1{z^3}e^{-4\kappa^2z^2}\partial_z\bigg)-t+\frac {S_j}{z^2}\bigg)\,G_0(j, t,  z, z')=z^3e^{4\kappa^2 z^2}\delta(z-z')
    \eea
\end{widetext}
with $t=K^2$ and $S_j=m_5^2R^2+m_j^2R^2$.

The two independent homogeneous regular and singular solutions $L_z\,y_{1,2}(z)=0$ to \eqref{KDIFF1} can be used to obtain
\bea
\label{GZZP}
G(z,z')=\frac 1{wp_0{\mathbb W}}y_1(z_<)y_2(z_>)
\eea
with the Wronskian ${\mathbb W}=(y_1'y_2-y_2'y_1)$. Note that the combination $wp_0{\mathbb W}$ is a constant independent of $z,z'$, and that 
(\ref{GZZP}) is symmetric in $z,z'$. This is more evident,
if we use the solutions to the eigenvalue problem
$L_zy_n=\lambda_ny_n$, with
\bea
G(z,z')=\sum_n\frac{y_n(z)y_n^*(z')}{\lambda_n}
\eea
with the normalizations
\bea
\int dz\,w(z)y_n^*(z)y_m(z)=\delta_{nm}
\eea
following from the hermiticity $L_z=L_z^\dagger$ in ${\mathbb R}$. 
To re-arrange the differential form (\ref{KDIFF1}), we 
redefine the spin-j propagator
\bea
G_0(j,t,z,z')\rightarrow (zz')^{\frac 32}
e^{\frac 34 \kappa^2(z^2+{z'}^2)}\, G_0(j, t, u, u')\nonumber\\
\eea
with $u=\kappa z$ and $\kappa^2=\frac 83 \kappa^2$,
which is now seen to solve 
\begin{widetext}
\bea
\label{apKDIFF2}
-\frac{d^2}{du^2}\,G_0+
\bigg(\frac{S_j+\frac {15}4}{u^2}
+\frac 94 u^2-\frac{ t}{\kappa^2}+3\bigg)\,G_0=\frac{e^{\frac{3}{4}(u^2-{u'}^2)}}{\kappa}\left(\frac{u}{u'}\right)^{3/2}\delta(u-u')
\eea    
Evaluating the right hand side at the delta function, the $u$ dependence is seen to drop out and we arrive at a standard Green's function problem.
Using the rescaling $u\rightarrow \sqrt{3}u$ and $\tilde t=t/3\kappa^2$, (\ref{apKDIFF2}) reads
\bea
\label{apKDIFF3}
-\frac{d^2}{du^2}\,G_0+
\bigg(\frac{S_j+\frac {15}4}{u^2}
+\frac{u^2}4-\tilde{t}+1\bigg)\,G_0=\frac{1}{\sqrt{3}\kappa}\delta(u-u')\nonumber\\
\eea
\end{widetext}
If we define $v=\frac 12 u^2$ and rescale the resulting propagator
\bea
\label{WHIT1}
G_0(j,t,v,v')=\frac 1{(vv')^{\frac 14}}\,
K_0(j,t,v,v')
\eea
(\ref{eq:SL}) can be mapped on the Whittaker
equation 
\bea
\label{WHIT2}
\frac{d^2K_0}{dv^2}+\bigg(\frac {\frac 14-\alpha^2}{v^2}+\frac \beta v-\frac 14 \bigg)\,K_0=-\frac{\delta(v-v')}{\sqrt{6}\kappa}\nonumber\\
\eea
with 
\bea
\alpha=\frac 12 (\Delta_g(j)-2)\qquad \beta=\frac 12(\tilde{t}-1)
\eea
The independent homogeneous solutions to (\ref{WHIT2}) are Whittaker functions
\bea
\label{WHIT3}
K_1(v)&=&e^{-\frac v2}v^{\frac 12+\alpha}\,
\mathbb M\bigg(\frac 12 +\alpha-\beta, 1+2\alpha, v\bigg)\nonumber\\
K_2(v)&=&e^{-\frac v2}v^{\frac 12+\alpha}\,
\mathbb U\bigg(\frac 12 +\alpha-\beta, 1+2\alpha, v\bigg)\nonumber\\
\eea
with Kummer  $\mathbb M$ (regular at $v=0$)  and Tricomi $\mathbb U$ (irregular with branch cut at $v=0$) hypergeometric functions. The inhomogeneous solution to (\ref{WHIT2}) is then
\bea
\label{apPROP1}
K_0(v, v')&=&\frac 12 {\cal A}\, K_2(v)K_1(v')\qquad v>v'\nonumber\\
K_0(v, v')&=&\frac 12 {\cal A}\, K_1(v)K_2(v')\qquad v<v'\nonumber\\
\eea
with the normalization fixed by the Wronskian
\bea
{\cal A}^{-1}=-\sqrt{6}\kappa{\cal W}(K_2,K_1)=
-\frac{4\kappa\Gamma(1+2\alpha)}{\Gamma\bigg(\frac 12 +\alpha-\beta\bigg)}
\nonumber\\
\eea
The confining bulk-to-bulk propagator is thus given by
\bea
\label{eq:b2bSWsingular}
 &&G_0(j, t, z, z')=\nonumber\\
 &&-(z z')^2(4\kappa^2zz')^{\Delta_g(j)-2}\frac{\Gamma(\frac{\Delta_g(j)-\tilde t}{2})}{\Gamma(\Delta_g(j)-1)}\mathbb{M}(z)\mathbb{U}(z')\nonumber,
\eea
where we introduced the shorthand $$\mathbb{M}(z)=\mathbb{M}(\frac{\Delta_g(j)-\tilde{t}}{2},\Delta_g(j)-1,4\kappa^2z^2)\,.$$ 


In order to recover the gravitational form factor obtained in \cite{Mamo:2022jhp}, we need to perform a Kummer transformation $$\mathbb U(a,b,z)=z^{1-b}\mathbb U(a-b+1,2-b,z)$$ and take the limit $z\to 0$. Which coincides with the bulk-to-boundary propagator quoted in \cite{Mamo:2022jhp} once the $z^2$ term from $G_0$ is taken into
account.

In light of the preceeding calculations to evaluate the Sommerfeld-Watson transform via saddle point we rewrite the Whittaker function involving the Tricomi $\mathbb U$ to better display its singular character. This is best seen
by rewriting it in terms of Kummer ${\mathbb M}$ functions, using the 
identity
\begin{widetext}
\bea
    K_2(v)=e^{-\frac v2}v^{\frac 12+\alpha}\left(\frac{\Gamma(-2\alpha)\mathbb M\bigg(\frac 12 +\alpha-\beta, 1+2\alpha, v\bigg)}{\Gamma(\frac{1}{2}-\alpha-\beta)}
+\frac{\Gamma(2\alpha)\mathbb M\bigg(\frac 12 -\alpha-\beta, 1-2\alpha, v\bigg)}{\Gamma(\frac{1}{2}+\alpha-\beta)}v^{-2\alpha}\right)
\eea
The singular part in $v$ is subleading in the saddle-point approximation and hence will be dropped in the following. Reverting the rescalings and coordinate transformations, we arrive again at the symmetric spin-j bulk-to-bulk propagator
\bea
\label{eq:b2bSWP}
 G_0(j, t, z, z')&=&-(z z')^2(4\kappa^2zz')^{\Delta_g(j)-2}\frac{\Gamma(\frac{\Delta_g(j)-\tilde t}{2})\Gamma(2-\Delta_g(j))}{\Gamma(\frac{4-\tilde t-\Delta_g(j)}{2})\Gamma(\Delta_g(j)-1)}\mathbb{M}(z)\mathbb{M}(z')\nonumber
\eea
where we introduced the shorthand
\bea
\mathbb{M}(z)=\mathbb{M}(\frac{\Delta_g(j)-\tilde{t}}{2},\Delta_g(j)-1,4\kappa^2z^2)\,.\nonumber
\eea

\subsection{Conformal limit}
In the confining case, the bulk-to-bulk propagator is given by
\bea
 G_0(j, t, z, z')&=&-(z z')^2(4\kappa^2zz')^{\Delta_g(j)-2}\frac{\Gamma(\frac{\Delta_g(j)-\tilde t}{2})\Gamma(2-\Delta_g(j))}{\Gamma(\frac{4-\tilde t-\Delta_g(j)}{2})\Gamma(\Delta_g(j)-1)}\mathbb{M}(a,b,4\kappa^2z^2)\mathbb{M}(a,b,4\kappa^2{z'}^2)\nonumber
\eea
where $a=\frac{\Delta_g(j)-\tilde{t}}{2}$ and $ b=\Delta_g(j)-1$.
As $\kappa\to 0$ we have $\tilde t=t/8\kappa^2\to\infty$ and we can rewrite 
\bea
\lim_{a\to\infty}\mathbb{M}({a,b,-x/a})=\Gamma(b)x^\frac{1-b}{2}J_{b-1}(2\sqrt{x})
\eea
where in our case $x=|t|z^2/4$. For small $\kappa$ we thus obtain
\bea
 G_0(j, t, z, z')&=&-(z z')^2(4\kappa^2zz')^{\Delta_g(j)-2}\frac{\Gamma(2-\Delta_g(j))}{\Gamma(\Delta_g(j)-1)}\left(\frac{|t| z z'}{2}\right)^{2-\Delta_g(j)}J_{\Delta-2}(\sqrt{|t|}z)J_{\Delta-2}(\sqrt{|t|}z')\nonumber
\eea
The Regge trajectory can again be resummed by means of a Sommerfeld-Watson transform
    \bea
G_2(s,t,z,z')=-(z z')^2\int\frac{dj}{4\pi i}&&\bigg(\frac{1+e^{-i\pi j}}{\sin\pi j}(\alpha'szz')^j(4\kappa^2zz')^{\Delta_g(j)-2}\nonumber\\
&&\times \frac{\Gamma(2-\Delta_g(j))}{\Gamma(\Delta_g(j)-1)}\left(\frac{|t| z z'}{2}\right)^{2-\Delta_g(j)}J_{\Delta-2}(\sqrt{|t|}z)J_{\Delta-2}(\sqrt{|t|}z')\bigg)
    \eea
    to obtain
\bea
{G_2(j,t,z,z')=-\frac{f^+(\lambda)}{2}\sqrt{\frac{\mathcal{D}}{4\pi\chi}}(z z')^2(\alpha' s z z')^{j_{\mathbb P}}e^{-\frac{\log (z z' |t|)^2}{4\mathcal{D}\tau}}}J_{0}(\sqrt{t}z)J_{0}(\sqrt{t}z').
\eea
\end{widetext}
In the Regge limit the Bessel functions become trivial at the saddle point and hence we recover the conformal result in \eqref{GCFt}.
It is interesting to note that the Regge limit must only be carried out after evaluating the Sommerfeld-Watson transform, otherwise the $t$ dependence would fully drop out as can be seen from \eqref{BESSX}. There is also an ambiguity in the sign of $\alpha$ in \eqref{WHIT2} and \eqref{SCALARX} since $\alpha=(\Delta-2)/2=\sqrt{S_j+4}/2$ and both the Whittaker as well as the Bessel equation are symmetric under $\alpha\to -\alpha$. This ambiguity also vanishes in the saddle point approximation and limit of small $\sqrt{\lambda}/\tau$ since the order of the Bessel function is then integer and we can use the reflection formulas to obtain the correct symmetry in $z$ and $z'$ as required by the Green's function of a self-adjoint operator. The Whittaker function $M_{\beta,\alpha}$ which is $K_1$ in our case is symmetric under $\alpha\to -\alpha$ for all $\alpha$.

\subsection{Mode sum}
 The mode decomposition for the spin-j bulk-to-bulk propagator is~\cite{Mamo:2022jhp}
\be
\label{KPOLES}
G_0(j, K,  z, z')=-\sum_n\frac{\psi_n(j,z)\psi_n(j, z')}{K^2+m_n^2(j)}
\ee
with the  wavefunctions $\psi_n(j,z)$  given by
\be
 \label{ZJV}
&&\psi_n(j,z)=c_n(j)\,z^{\Delta_{g}(j)}L_{n}^{\Delta_{g}(j)-2}(4\kappa^2 z^2)\,,\nonumber\\
\ee
with
\bea
\Delta_g(j)=2+\sqrt{2\sqrt{\lambda}(j-j_{\mathbb P})}
\eea
and normalized by
\be
c_n(j)=\left(\frac{2(4\kappa^2)^{\Delta_q(j)-1}\Gamma(n+1)}{\Gamma(n+\Delta_q(j)-1)}\right)^{\frac 12}.
\ee
The squared mass spectrum is
\be
m^2_n(j)=16\kappa^2\bigg(n+\frac 12 \Delta_g(j)\bigg).
\ee
In this casting,  the role played by the Regge
poles is transparent. However, the Reggeization requires summing over the full Regge  trajectory.

Starting from the Sommerfeld-Watson transform of the bulk-to-bulk propagator in impact parameter space we have
\begin{widetext}
\bea
\label{GOCONF}
G_1(s,t,z,z')=\int\,d^2b_\perp e^{-iqb_\perp}\,
\int_{n_L} \frac{dj}{4\pi i}
\bigg(\frac{1-e^{-i\pi (j-1)}}{{\rm sin}(\pi (j-1))}\bigg)\, 
(\alpha'szz')^{j-1}\,\tilde G_0(j, b,  z, z'),
\eea
\end{widetext}
where the scalar bulk-to-bulk propagator in impact parameter space is given by
\bea
\label{KPOLESb}
\tilde G_0(j, b,  z, z')=
\sum_n\psi_n(j,z)\psi_n(j, z')\,\frac {K_0(m_n(j)b)}{2\pi zz'},\nonumber\\
\eea
obtained by the Fourier transform
\bea
\int d^2q \frac{e^{iqb_\perp}}{q^2+m_n(j)^2}=\frac{K_0(m_n(j)b)}{2\pi}.
\eea
The  wavefunctions follow from (\ref{ZJV}) for odd spin, with the Reggeized odd spin glueball mass
spectrum
\bea
m^2_n(j)&=&(4\kappa)^2(n+\frac 12 \Delta_g(j))\nonumber\\
&\equiv& m_0^2\bigg(n+1+\frac 12 \sqrt{2\sqrt{\lambda}(j-j_{\mathbb O})}\bigg).\nonumber\\
\eea
The dominant contribution in (\ref{GOCONF}) stems from the large $b$ asymptotic of $K_0$ 
where
\be
K_0(m_n(j)b)\approx e^{-m_n(j)b}\sqrt{\frac{\pi}{2m_n(j)b}}.
\ee
Due to the soft wall, the bulk wavefunctions will be localized at small $z$. To consider the Laguerre polynomials in this limit we can write
\bea
\label{ZLAG}
z^{\Delta}L_n^{\Delta-2}(z^2)&=&\frac{z^{-2}e^{z^2}}{\Gamma(n+1)}\int dxe^{-x} x^{n+\alpha/2}J_{\Delta-2}(2\sqrt{xz})\nonumber\\
&\approx &\frac{z^\Delta}{\Gamma(n+1)\Gamma(\Delta-1)}\int dxe^{x}x^{n+\Delta-2}\nonumber\\
&=&z^\Delta \frac{\Gamma(n+\Delta-1)}{\Gamma(\Delta-1)\Gamma(n+1)}.
\eea
From the discussion of the conformal limit, we know that the Gamma functions will thus give a subleading contribution in the saddle point approximation. Continuing with the evaluation of the Odderon propagator we now have
\begin{widetext}
\bea
G_1(s,t,z,z')\approx \int\,d^2b_\perp e^{-iqb_\perp}\,
\int_{C_L} \frac{dj}{4\pi i}
\bigg(\frac{1-e^{-i\pi (j-1)}}{{\rm sin}(\pi (j-1))}\bigg)\, 
(\alpha'szz')^{j-1}\,
\sum_n\psi_n(j,z)\psi_n(j, z')\,\frac{e^{-m_n(j)b}}{2\pi zz'}\bigg(\frac \pi{2m_n(j)b}\bigg)^{\frac 12}\nonumber\\
\eea
The contour $C_L$ in (\ref{GOCONF}) is defined to the left of the branch point $j=j_{\mathbb O}$ as in Fig.~\ref{fig:cuts}. In light of the arguments in the main text and for $\tau/\sqrt\lambda\gg 1$, we carry the j-integration along $C_L$ by saddle point, with the result
 \bea
 \label{G1XX}
G_1(s,t,z,z')\approx-f^-(\lambda)\int\,d^2b_\perp e^{-iqb_\perp}\frac{1}{8}\sqrt{\frac{m_0b}{32\pi^2(n+1)\mathcal{D}\tau^3}}\,\sum_n\frac{\psi_n(j_\mathbb{O},z)\psi_n(j_\mathbb{O}, z')}{zz'}e^{(j_\mathbb{O}-1)\tau-m_0b\sqrt{n+1}-\frac{(m_0b)^2}{64(n+1)\mathcal{D}\tau}}.\nonumber\\
\eea   
\end{widetext}

\section{Conformal limit}
\label{CFTX}
In the Regge limit $s\gg -t$,
 the conformal limit, $\kappa\to 0$, is best sought by noting that \eqref{eq:SL} reduces to
\bea
\label{SCALARX}
-\frac {d^2}{dz^2}\,G_0+\bigg(\frac{S_j+\frac {15}4}{z^2}-t\bigg) \,G_0(z)=\delta(z-z'),\nonumber\\
\eea
where we employed the rescaling $$G_0(j,t,z,z')\to(z z')^{3/2}G_0(j,t,z,z').$$ The two independent homogeneous solutions are Bessel functions $J$ (regular at the origin) and $Y$ (singular at the origin)
\bea
\label{BESS}
G_1(z)&=&\sqrt z\,J_{-\sqrt{S_j+4}}(\sqrt t z)\nonumber\\
G_2(z)&=&\sqrt z\,Y_{-\sqrt{S_j+4}}(\sqrt t z)
\eea
and hence
\bea
G_0(j,t,z,z')&=&\mathcal{A}(z z')^{2}J_{-\nu}(\sqrt t z_<),Y_{-\nu}(\sqrt t z_>)\nonumber\\
\nu&=&\Delta_g(j)-2
\eea
with the normalization fixed by the Wronskian
\be
\mathcal{A}^{-1}=\mathcal{W}(G_1(z),G_2(z))=\frac{2}{\pi}.
\ee
After reverting the rescaling, in the AdS limit, the bulk-to-bulk scalar propagator is 
\bea
&&G_0(j,t,z,z')=\nonumber\\
&&\frac{\pi } 2(z z')^{j_\pm}\,
J_{2-\Delta_g(j)}(\sqrt{t} z_<)\,Y_{2-\Delta_g(j)}(\sqrt{t} z_>).\nonumber\\
\eea
In the Regge limit we also have  $x=\sqrt t\,z\ll 1$, for which the Bessel functions simplify
\bea
\label{BESSX}
J_{-\nu}(x)&\approx& \frac 1{\Gamma(1-\nu)}\bigg(\frac x2\bigg)^{-\nu}\nonumber\\
Y_{-\nu}(x)&\approx& -\frac{1}{\pi}{\rm cos}(\nu \pi)\,\Gamma(\nu)\,\bigg(\frac x2\bigg)^{-\nu}
\eea
With this in mind, the Sommerfeld-Watson transform \eqref{SWT} becomes
\begin{widetext}
    \bea
    \label{GpmCF}
G_{j_\pm}(s,t,z,z')=-\int\frac{dj}{4\pi i}\frac{1+ e^{-i\pi (j-j_\pm)}}{\sin\pi (j-j_\pm)}\frac{\cos\pi(\Delta_g(j)-2)}{\Gamma(3-\Delta_g(j))}\Gamma(2-\Delta_g(j))\left(\frac{tzz'}{4}\right)^{2-\Delta_g(j)}(\alpha'szz')^{j-j_\pm}\nonumber\\
    \eea
We will evaluate \eqref{GpmCF} again via saddle point approximation. The branch cut of $\Gamma(2-\Delta_g(j))=\Gamma(iy)\approx e^{-i \gamma_E y}/iy$ at $\Delta_g(j)-2=-iy=$ is chosen to the left of the integration contour, along the negative real axis. In the large $s/|t|$ limit, the integral is dominated by the saddle point
$$\Delta_g(j)-2=\left(\frac{{\rm log}(zz'|t|)}{2{\cal D}\chi}\right)^2\rightarrow 0$$
with the result 
\bea
\label{GCFt}
{G_{j_\pm}(j,t,z,z')=f^{\pm}(\lambda)\sqrt{\frac{\mathcal{D}}{4\pi\chi}}\frac{(z z')^{j_\pm}}{2}(\alpha' s z z')^{j_{\mathbb{P/O}}-j_\pm}e^{-\frac{\log (z z' |t|)^2}{4\mathcal{D}\chi}}}
\eea
\end{widetext}
for fixed but large rapidity $\chi$, with $j_{\mathbb{P/O}}$ given by \eqref{eq:jPconf}, \eqref{eq:jOconf} and where 
\be
f^+(\lambda)=i+\frac{\sqrt{\lambda}}{\pi}-\frac{\pi}{3\sqrt{\lambda}}
\label{eq:ConfEvenSig}
\ee
and
\be
f^-(\lambda)=i+\frac{4\sqrt{\lambda}}{\pi m_k^2}-\frac{m_k^2\pi}{12\sqrt{\lambda}}\,.
\label{eq:ConfOddSig}
\ee
The signature factor for the $k=2$ branch again coincides with the Pomeron signature factor. As is the case for the repulsive wall, due to the pole the intercept cannot be fixed at 1 with $m_k^2=0$. In the conformal limit the diffusion is logarithmic in
$z$, with the latter identified with the size of the transverse dipoles or string bits composing the exchanged Pomeron/Odderon.

\subsection{Alternative derivation: forward region}

Alternatively and in  the conformal limit with $\kappa\to 0$ and  $t=0$, the ensuing scalar propagator is a superposition of conformal plane waves 
\bea
\label{G0CF}
G_0(j,0,z,z')=
\int \frac{d\nu}{2\pi}\frac{e^{i\nu(\rho-\rho')}}{4\nu^2+4+m_\pm^2}
\eea
with $z=e^{-\frac \rho 2}$ and $m_+^2=m_j^2R^2,\ m_-^2=m_k^2R^2+m_j^2R^2$. Inserting \eqref{G0CF} in \eqref{SWT}
gives
\begin{widetext}
  \bea
\label{G0t}
G_{j_\pm}(s,0,z,z')=\sqrt{\frac{ {\cal D}}{4\pi\chi}}(zz')^{j_\pm}(\alpha's zz')^{j_{{\mathbb O}/\mathbb{P}}-j_\pm}\,
\,e^{-\frac{(\rho-\rho')^2}{4{\cal D}\chi}},
\eea  
\end{widetext}
with the shifted Pomeron, $j_\mathbb{P}$, and Odderon, $j_{\mathbb{O}}$, intercepts.
The shift is caused by the diffusion in rapidity
$$\tau={\rm log}(\alpha' s{zz'})\sim\chi=\log(\alpha's)$$
in warped AdS, with a diffusion constant
 ${\cal D}=\frac 1{2\sqrt\lambda}$, in agreement with the original analysis in~\cite{Brower:2006ea,Brower:2008cy}.

\subsection{Alternative derivation: Finite impact parameter}
The forward bulk-to-bulk part propagator can also be obtained from its explicit form for a finite impact parameter. For that, 
consider the bulk equation
\begin{widetext}
\bea
\label{SCALAR3}
\bigg(-\frac {1}{\sqrt  {-g}}\partial_z\,g^{zz}\sqrt{-g}\,\partial_z-z^2\partial_\perp^2+S_j\bigg)\,\tilde G_0(j,b,z,b',z')=\frac{\delta(z-z')}{\sqrt{-g}}\delta^2(b_\perp-b'_\perp)
\eea
which is the Fourier inverse of \eqref{SCALARX}. For simplicity, we will restrict our discussion to the Odderon propagator,  the Pomeron propagator follows similarly. The solution to \eqref{SCALAR3} is
\bea
\tilde G_0(j,b,z,b',z')=
\frac{1}{4\pi zz'}\,\frac{e^{-\sqrt{S_j}\xi}}{{\rm sinh}\xi}
\eea
with $\xi$ fixed by the chordal distance in AdS
\bea
{\rm cosh}\xi=1+\frac{(z-z')^2+(b-b')^2}{2zz'}
\eea
We note that $S_j=2\sqrt\lambda (j-j_{\mathbb O})$ develops a branch point at $j=j_{\mathbb O}$ as seen in Fig.~\ref{fig:cuts}. 
Using the Sommerfeld-Watson formula \eqref{SWT}, we have 
\bea
\label{eq:SWTb}
G_1(s,t,z,z')=\int d^2b_\perp\,e^{iq\cdot b_\perp}
\int \frac{dj}{4\pi i}\bigg(\frac{s^{j-1}+(-s)^{j-1}}{{\rm sin}(\pi (j-1))}\bigg)\, (\alpha'zz')^{j-1}\,\tilde G_0(j,b_\perp,z, b'_\perp, z')
\eea
with the contour integral to the right of the branch point $j_{\mathbb O}$, and summing over the odd poles $j=1,3,...$. In the forward limit with  $t=q^2=0$, we can switch 
$b_\perp \rightarrow \xi$ with the measure
$$d^2b_\perp=2\pi zz'{\rm sinh}\xi\,d\xi$$
and deform the contour to the left along the cut $C_L$, to obtain
\bea
G_1(s,0,z,z')=\int_{\xi_0}^\infty\, 2\pi zz'd\xi\,
\int_{C_L} \frac{dj}{4\pi i}
\bigg(\frac{1-e^{-i\pi (j-1)}}{{\rm sin}(\pi (j-1))}\bigg)\, 
(\alpha'szz')^{j-1}\,
\frac{e^{(2-\Delta_g(j))\xi}}{4\pi}
\eea
with $\xi_0=|{\rm log}z/z'|$.
In the double limit of large rapidities $\tau={\rm log}\big(\alpha'szz'\big)\gg 1$ and strong coupling $\sqrt\lambda\gg 1$, the j-integration can be evaluated 
in leading order in $\sqrt\lambda/\tau\ll 1$, following
the original arguments in~\cite{Brower:2007xg}, with the result
\bea
\label{ODD00}
G_1(s,0,z,z')\approx \frac{zz'}{4}\,
\big(\alpha'szz'\big)^{j_{\mathbb O}-1}
f^-(\lambda)\bigg(\frac{\sqrt\lambda}{2\pi}\bigg)^{\frac 12}\,
\int_{\xi_0}^\infty d\xi\,\xi\,\frac{e^{-\frac{\xi^2}{4D\tau}}}{\tau^{\frac 32}}
\eea
\end{widetext}
We  can unwind the remaining Gaussian integral, to have 
\bea
\label{ODD00X}
G_1(s,0,z,z')\approx 
f^-(\lambda)\,
\,\big(\alpha'szz'\big)^{j_{\mathbb O}-1}
\,\frac{z z'}{2}\sqrt{\frac{\mathcal{D}}{4\pi \tau}}\,e^{-\frac{\xi_0^2}{4D\tau}}\nonumber\\
\eea
and similarly for the Pomeron
\bea
\label{EVEN00X}
&&G_2(s,0,z,z')\approx \nonumber\\
&&f^+(\lambda)\,\big(\alpha'szz'\big)^{j_{\mathbb P}-2}
\,\frac{(z z')^2}{2}\sqrt{\frac{\mathcal{D}}{4\pi \tau}}\,e^{-\frac{\xi_0^2}{4D\tau}}\nonumber\\
\eea
in agreement with (\ref{G0t}).

\section{Bulk Dirac fields with a repulsive wall}
\label{ap:BulkDiracRWall}
To construct the full $pp$ and $p\bar p$ scattering amplitudes,
we identify the proton (antiproton) with bulk Dirac fermions. The bulk nucleon as a Dirac field is described by the chiral pair  $\Psi_{1,2}$, with $1,2$ referring to the boundary chirality  $1,2=\pm=R,L$.  They are dual to the boundary sources $\Psi_{\pm}\leftrightarrow{\cal O}_\pm$ with anomalous dimension $\pm M=\pm(\Delta-2)=\pm(\tau-3/2)$.  For that, consider the action of a free Dirac fermion in 5D curved space
\begin{widetext}
    \bea
        S_F=\frac{1}{g_5^2}\int d^5x \sqrt{g}\left(\frac{i}{2}\overline{\Psi}_{1,2}e^M_a\Gamma^a\overleftrightarrow{D}_M\Psi_{1,2}-\left(\pm M\right)\overline{\Psi}_{1,2}\Psi_{1,2}\right),\nonumber\\
        \label{eq:BulkDiracFree}
    \eea
with $$\frac 12 {\rm ln}{g(z)}=2A(z)=2\log\frac{R}{z}+a\kappa^2z^2$$ and the covariant derivative given by $$ D_M=\partial_M+\frac{1}{8}\omega_M^{\ AB}\left[\Gamma_A,\Gamma_A\right].$$ More specifically we have
\be
    S_F=\frac{1}{g_5^2}\int d^4xdz e^{4A(z)}\left\{-\frac{i}{2}\overline{\Psi}_{1,2}\left[\slashed{\partial}-2i\gamma_5 A'(z)-i\gamma_5\partial_z\right]\Psi_{1,2}+e^{A(z)}M\overline{\Psi}_{1,2}\Psi_{1,2}\right\}
\ee
which, upon the field redefinition $\Psi\to g_5e^{-2A(z)}\Psi$, reduces to
\be
    S_F=\int d^4xdz\left\{-\frac{i}{2}\overline{\Psi}_{1,2}\left[\slashed{\partial}-i\gamma_5\partial_z\right]\Psi_{1,2}+e^{A(z)}M\overline{\Psi}_{1,2}\Psi_{1,2}+h.c.\right\}.
\ee
The equation of motion is thus given by
\be
\left[-i\slashed{\partial}-\gamma_5\partial_z+Me^{A(z)}\right]\Psi=0.
\ee
Performing a chiral Kaluza-Klein decomposition as
\bea
\Psi_1(p,z;n)&=&\psi_R(z;n)\Psi^0_{R}(p)+ \psi_L(z;n)\Psi^0_{L}(p)\nonumber\\
\Psi_2(p,z;n)&=&\psi_R(z;n)\Psi^0_{L}(p)+ \psi_L(z;n)\Psi^0_{R}(p)\nonumber\\
\Psi_{L/R}^0&=&\sum_ne^{ik_n x}\frac{1\mp\gamma_5}{2}u_s(k)f_{L/R}(z)
\eea
we obtain the coupled equation of motion
\bea
\left(\partial_z\pm M e^{A(z)}\right)f_{L/R}=\pm m_n f_{R/L}
\eea
which can be decoupled by iteration to give
\bea
    \left(\partial_z^2\pm MA'(z)e^{A(z)}-M^2e^{2A(z)}\right)f_{L/R}(z)= m_n^2 f_{L/R}(z).
\eea    
Linearizing the dilaton contribution $$e^{2A(z)}\approx \left(\frac{R}{z}\right)^2\left(1+a\kappa_N^2z^2\right),$$ we obtain
\be
\label{eq:LinDirac}
    \left(-\partial_z^2+\frac{M(M\pm 1)}{z^2}+ a^2\kappa_N^4z^2-\tilde m_n^2 \right)f_{L/R}(z)=0,
\ee  
where $\tilde{m}_n^2=m_n^2+M(M\pm 1)a\kappa_N^2$ and we did not expand $e^{A(z)}$ to obtain a chirally symmetric mass spectrum. Note the introduction of a new mass scale, $\kappa_N$, to account for the different Regge trajectories of the baryon and the glueball spectrum.
Upon performing the coordinate transformation $u=a\kappa_N^2z^2$ and a further field redefinition of $f_{L/R}(u)= e^{-\frac{u}{2}}u^{\frac{1+\sqrt{1+4M(M\pm 1)}}{4}}f_{L/R}(u)$ we arrive at the equation
    \be
uf_{L/R}''(u)+(\alpha+1-u)f_{L/R}'(u)+n f_{L/R}(u)=0,\nonumber\\
    \ee
where
\bea
\alpha&=&\frac{1}{2}\sqrt{1+4M(M\pm 1)}\nonumber\\
n&=&\frac{\tilde{m}_n^2}{4a\kappa_N^2}-\frac{2+\sqrt{1+4M(M\pm 1)}}{4}.
\eea
This is the Sturm-Liouville normal form for the associated Laguerre polynomials $L_n^\alpha(u)$. Hence the eigenvalues are given by
  $$m_n^2=4a\kappa_N^2\left(n+\frac{1}{2}-\frac{M(M\pm 1/2)+\sqrt{1+4M(M\pm 1)}}{4}\right)$$  
\end{widetext}
For fermions with positive parity we have $M=\Delta-\frac{d}{2},\ \tau=3$ and $\Delta=\tau+\frac{1}{2}$ giving $M=1$ for $d=5$ and hence

\bea
 m_n^2&=&4a\kappa_N^2\left(n+\frac{3}{4}\right)\nonumber\\
 \alpha&=&\frac{2\pm 1}{2},
\eea

Reverting the rescalings we thus obtain
\bea
\label{eq:DiracWFLin}
\psi_L(n,z)=n_L a\kappa_N^2z^4e^{-\frac{3}{2}a\kappa_N^2z^2} L_n^{3/2}(a\kappa_N^2z^2)\nonumber\\
\psi_R(n,z)=n_R a\kappa_N^2z^4e^{-\frac{3}{2}a\kappa_N^2z^2} L_n^{1/2}(a\kappa_N^2z^2)
\eea

The normalization is fixed by the condition
\bea
\int dz n_{L/R} {n_{L/R}}^*f_{L/R}^n(z)f_{L/R}^m(z)=\delta_{mn}\nonumber\\
\eea

which gives
\bea
n_L&=&\sqrt{\frac{2\sqrt{a}\kappa n!}{\Gamma(n+5/2)}}\nonumber\\
n_R&=&n_L^n\sqrt{n+3/2}.
\eea

From the mass eigenvalues it is apparent that one needs to use the positive linearized dilaton in order to not obtain a tachyonic solution. We will refer to this background as repulsive wall. As is the case for the bulk-to-bulk propagator, the fermionic spectrum still reggeizes properly in the repulsive wall model. When the spectrum is matched to the proton mass, the comparision is less desirable, resulting in an increase of the coupling by more than 60\% compared to the soft wall.

The non-normalizable modes for the bulk Dirac fields are given in terms of Kummer functions
\bea
\tilde\psi_R(n, z)&=&N_R\,U\bigg(-n, 3/2, a\kappa_N^2z^2\bigg)\nonumber\\
\tilde\psi_L(n, z)&=&N_L\,U\bigg(-n, 1/2, a\kappa_N^2z^2\bigg),
\eea
which can be recast as a sum over Regge poles
\bea
\tilde\psi_R(p, z)&=&\sum_{n=0}^\infty\frac{f_np\tilde\psi_R(n,z)}{p^2-m^2}\nonumber\\
\tilde\psi_L(p, z)&=&\sum_{n=0}^\infty\frac{f_n m_n\tilde\psi_R(n,z)}{p^2-m^2},
\eea
with $f_n=\kappa_N/n_R$. When calculating the amplitude, the LSZ reduction will pick up the residue of the corresponding pole, effectively reducing the bulk to boundary propagator for the proton to
\be
\tilde{\psi}_{L/R}(p,z)= f_0\ m_p \psi_{L/R}(0,z)
\ee
Further, in the Regge limit, we can utilize 
\be
\overline{u}(p_2)\gamma^\mu u(p_1)=\overline{v}(p_2)\gamma^\mu v_(p_1)=(p_1+p_2)^\mu\delta_{s_1s_2}\nonumber\\
\ee
where we introduced the shorthand $u(p_i)=u_{s_i}(p_i)$ and analogously for $v$. 



\bibliography{TOTEM}

\end{document}